\documentclass[debug]{rmaa}

\title{Abundances in photoionized nebulae of the Local Group and nucleosynthesis 
of intermediate mass stars}

\author{
  W. J. Maciel,\altaffilmark{1} 
  R. D. D. Costa,\altaffilmark{1}
  and O. Cavichia\altaffilmark{2}}

\altaffiltext{1}{Instituto de Astronomia, Geof\'\i sica e Ci\^encias
                 Atmosf\'ericas, Universidade de S\~ao Paulo, Brazil.}
\altaffiltext{2}{Instituto de F\'\i sica e Qu\'\i mica, Universidade 
                 Federal de Itajub\'a, Brazil.}

\shortauthor{Maciel, Costa \& Cavichia}
\shorttitle{Abundances in photoionized nebulae}

\fulladdresses{
\item W. J. Maciel, and R. D. D. Costa: 
    Instituto de Astronomia, Geof\'\i sica e Ci\^encias
    Atmosf\'ericas, Universidade de S\~ao Paulo - 
    Rua do Mat\~ao 1226, CEP 05508-090, S\~ao Paulo SP, Brazil
    (wjmaciel@iag.usp.br, roberto.costa@iag.usp.br.)

\item O. Cavichia: 
     Instituto de F\'\i sica e Qu\'\i mica, Universidade Federal 
     de Itajub\'a, Av. BPS 1303, Pinheirinho, CEP 37500-903, 
     Itajub\'a, MG, Brazil 
    (cavichia@unifei.edu.br.)}

\listofauthors{W. J. Maciel, R. D. D. Costa & O. Cavichia}
\indexauthor{Maciel, W. J.}
\indexauthor{Costa, R. D. D.}
\indexauthor{Cavichia, O.}

\abstract{
Photoionized nebulae, comprising HII regions and planetary nebulae, are excellent laboratories 
to investigate the nucleosynthesis and chemical evolution of several elements in the Galaxy 
and other galaxies of the Local Group. Our purpose in this investigation is threefold: 
(i) compare the abundances of HII regions and planetary nebulae in each system in order 
to investigate the differences derived from the age and origin of these objects, (ii) compare 
the chemical evolution in different systems, such as the Milky Way, the Magellanic Clouds, 
and other galaxies of the Local Group, and (iii) investigate to what extent the nucleosynthesis 
contributions from the progenitor stars affect the observed abundances in planetary nebulae, 
which constrains the nucleosynthesis of intermediate mass stars. We show that all objects in
the samples present similar trends concerning distance-independent correlations, and some
constraints can be defined on the production of He and N by the PN progenitor stars.
}

\resumen{ }

\addkeyword{Galaxies: Abundances}
\addkeyword{Galaxies: Disk}
\addkeyword{ISM: abundances}
\addkeyword{ISM: planetary nebulae: general}

\begin{document}
\maketitle

\section{Introduction}
\label{section1}

Planetary nebulae (PN) have strong emission lines of H, He, O, N, Ne, S, and Ar, including forbidden 
lines and recombination lines. The analysis of these lines gives abundances accurate to about 0.2 to
0.3 dex.  The abundances include elements that are probably not significantly produced by the progenitor 
stars (O, S, Ne, Ar), and therefore contribute to the study of the chemical evolution of the host galaxies. 
The abundances of these  elements in planetary nebulae are not expected to be affected in a signficant 
way by the evolution of their progenitor stars. In fact, the average abundances of these elements are not 
very different from the observed abundances in HII regions. The same is true for the correlations of the 
ratios Ne/H, Ne/O, etc. with the metallicity, as measured by the O/H ratio.  Therefore, the measured 
abundances of these elements in PN reflect the interstellar abundances at the time the progenitor stars 
were formed, and can be compared with the abundances of red giants, AGB stars, and  younger objects, such 
as HII regions and Blue Compact Galaxies (BCG), provided the age-metalliciticy relation and the presence 
of abundance gradients, both radial and vertical, are taken into account. 

Planetary nebulae also provide accurate abundances of some elements that are produced by the progenitor 
stars (He, N, C), so that  the analysis of these abundances and their distance independent correlations 
can be used to test nucleosynthesis predictions of theoretical models for intermediate mass stars. 

Some of these elements are difficult to study in stars, and are better observed  in photoionized nebulae. 
In fact, in recent years there has been an important advance in the determination of accurate abundances 
in PN and in HII regions, both in the Milky Way as well as in several external galaxies, particularly 
in the Local Group. Blue Compact Galaxies (BCG) and Emission Line Galaxies (ELG) can also be included 
as low metallicity HII regions, as we will see in the following sections. In this work, we consider the 
abundances of several chemical elements in PN based on our own results and on some recent data from the 
literature, and compare these results with the abundances of HII regions, Blue Compact Galaxies, 
and Emission Line Galaxies both in the Galaxy and in other objects of the Local Group. Our purpose
in this investigation is threefold: (i) compare the abundances of HII regions and PN in each system in 
order to investigate the differences derived from the age and origin of these objects, (ii) compare the 
chemical evolution in different systems, such as the Milky Way, the Magellanic Clouds, and other galaxies 
of the Local Group, and (iii) investigate to what extent the nucleosynthesis contributions from the 
progenitor stars affect the observed abundances in planetary nebulae, which constrains the 
nucleosynthesis of intermediate mass stars.

\section{The Data}
\label{section2}

The data used in this investigation include a large sample of PN and HII regions in the following 
galaxies: The Milky Way (MW), the Large Magellanic Cloud (LMC), the Small Magellanic Cloud (SMC), M31, 
M32, M33, M51, M81, M101, NGC 185, NGC 205, NGC 300, NGC 628, NGC 3109, NGC 5194, and the Sextans galaxy.
Typical uncertainties in the abundances are 0.2 to 0.3 dex for PN and 0.1 to 0.2 dex for HII regions,
especially for the elements with larger abundances, such as oxygen and nitrogen.
The total PN sample includes over 1300 objects, while the HII region sample has over 900 objects, so that 
a reasonably large sample of Local Group objects is considered, amounting to over 2200 photoionized nebulae.
Although not a complete sample, it is large enough to compensate for the inevitable degree of inhomogeneity,
so that some interesting conclusions on the chemical evolution of the systems considered can be obtained. 
The data sources are partially listed by Maciel et al. (\citeyear{maciel2014}), with the addition of some 
recent data as discussed below.

New data on the chemical abundances of planetary nebulae (PN) in different systems have been obtained by 
our group, hereafter referred to as the IAG sample, with important consequences on the nucleosynthesis of 
the PN progenitor stars and on the chemical evolution of  the Galaxy (see for example Idiart et al. 
\citeyear{idiart2007}, Costa et al. \citeyear{costa2008}, Maciel et al. \citeyear{maciel2009}, 
\citeyear{maciel2010a}, \citeyear{maciel2010b}, Maciel \& Costa \citeyear{mc2009}, Cavichia et al. 
\citeyear{cavichia2010}, \citeyear{cavichia2011}, \citeyear{cavichia2017}). These systems include the solar 
neighbourhood, the galactic disk, the galactic bulge and the Magellanic Clouds. Abundances of the main chemical 
elements have been determined, namely, He, N, O, Ne, S and Ar, and the derived abundances are expected to 
be corrected to 0.2 to 0.3 dex in average. Most PN abundances considered here have been derived on the
basis of the analysis of forbidden lines with the use of ionization correction factors (ICFs), and
the reader is referred to the papers above for details on the adopted procedures.

Apart from our own data, we have considered some recent abundance determinations from the literature,
particularly from sources using a similar procedure than our group, so that a direct comparison can
be made for the objects in common.

Girard et al. (\citeyear{girard}) analyzed a sample of 48 galactic PN with [WR] central stars, comprising 
[WC], [WO] and wels (weak emission line stars). In agreement with their own conclusions, we find no 
important differences between these objects and the \lq\lq normal\rq\rq\ PN, at least concerning the 
chemical abundances of the considered elements.

In a series of papers, Henry and collaborators (\citeyear{henry2000}, \citeyear{henry2004}, 
\citeyear{henry2010}, \citeyear{henry2012}) and Milingo et al. (\citeyear{milingo}), established accurate 
abundances of a large sample of galactic PN, comprising objects in the galactic disk and anticentre, 
forming an extremely homogeneous sample. Similar methods have been used to determine the abundances as 
in our own data, so that it is possible to compare the results of our group with these data, especially 
in view of our intended comparison with Local Group objects. Particularly important here is the discussion on the 
\lq\lq sulphur anomaly\rq\rq, as we will see in detail in the following section. We have also included 
the results of the very recent surveys by Garc\'\i a-Hern\'andez \& G\'orny (\citeyear{garcia-gorny}) and 
Delgado-Inglada et al. (\citeyear{delgado2015}), based on Spitzer data. Comparison of infrared data with 
optical spectroscopic data produce a particularly accurate set of abundances, as also discussed by 
Garc\'\i a-Hern\'andez et al. (\citeyear{garcia2016}). These investigations (see also Delgado-Inglada et 
al. \citeyear{delgado2014}) have considerably improved the determination of the ICFs, which are an essential 
part of the abundance determinations in PN from forbidden lines. The uncertainties are then in principle 
smaller than in previous investigations, but in fact the derived abundances are comparable, so that 
we feel safe to include these results in the present investigation.

Apart from our own data on the Magellanic Clouds, we have included samples from Stasi\'nska et al. 
(\citeyear{srm1998}, SRM) and Leisy \& Dennefeld (\citeyear{leisy}, LD). In agreement with some conclusions 
by Henry et al. (\citeyear{henry2010}), part of the results by Leisy \& Dennefeld  (\citeyear{leisy}) are 
upper limits or  overestimates, and have not been included in our sample. Our results can then 
be considered as an update and improvement over the results by Stasi\'nska et al. (\citeyear{srm1998}), 
where an attempt was made to compare the PN population in five galaxies: Milky Way, LMC, SMC, M31, and M32.  

Abundances of a large sample of PN in the Galaxy (disk, bulge, and halo) and in the Magellanic Clouds were 
analyzed by Milanova \& Kholtygin (\citeyear{milanova}). The similarity of the elemental abundances of 
these objects was pointed out, in agreement with our present results, taking into account the lower 
metallicities of the LMC and SMC. The average uncertainties in these abundance determinations are similar 
as in our data, typically of 0.1 dex for helium and 0.2 dex for the other elements. Finally, we have also 
used the compilation of PN data by Chiappini et al. (\citeyear{cgsb}), who included objects in the galactic 
bulge and disk, as well as in the Large Magellanic Cloud. This is a particularly large sample, so that we 
could use it as a check of the consisitency of the abundances of the sources considered, in order
to minimize the effects of the inhomogeneity of these sources.

The comparison samples of HII regions for the Milky Way and Magellanic Clouds are from Afflerbach et al. 
(\citeyear{afflerbach}), Guseva et al. (\citeyear{guseva}), Rudolph et al. (\citeyear{rudolph}), Tsamis 
et al. (\citeyear{tsamis}), and Delgado-Inglada et al. (\citeyear{delgado2015}). We have preferably adopted
abundances determined from detailed electron temperatures, which are usually more accurate than the
results from the strong line method. Some recent results by Reyes et al. (\citeyear{reyes}) for both 
clouds have also been taken into account. Additional data for HII regions in the Magellanic Clouds come 
from Peimbert et al. (\citeyear{peimbert}), who presented chemical abundances of the HII region NGC 346 
in the SMC based on spectrophotometric data obtained at the CTIO 4m telescope. We have here adopted the 
results for the preferred value of the fluctuation temperature parameter $t^2$, as given in their table 9. 
Also, Pe\~na-Guerrero et al. (\citeyear{pena-guerrero}) presented chemical abundances of two HII regions 
in the SMC, NGC 456 and NGC 460, based on long-slit spectrophotometry taking into account the presence of 
thermal inhomogeneities. We have adopted here the average of the positions reported in the paper. 
Selier \& Heydari-Malayeri (\citeyear{selier}) presented chemical abundances of He, O, N, and Ne for two 
outer HII regions in the Magellanic Clouds, LMC N191 and SMC N77. The data come from optical imaging and 
spectroscopic ESO NTT observations along with archive data. Finally, by combining optical and infrared data, 
Vermeij \& van der Hulst (\citeyear{vermeij})  determined element abundances for a sample of HII regions in 
the Large and Small Magellanic Cloud. We have considered 15 HII regions from their sample, with abundances 
of He, O, N, Ne, S, and Ar.

We have also included data on PN and HII regions in several external galaxies apart from the Magellanic 
Clouds, which is interesting in order to compare with the local data. Of course, as pointed out by Richer 
and McCall (\citeyear{richer2016}), PN that are far away are selected among the brightest objects, so 
that it remains to be investigated whether or not these objects are representative of the PN population.

Kwitter et al. (\citeyear{kwitter}) analyzed a sample of 16 PN in the outer disk of M31, for which they 
could derive electron temperatures and abundances. As we will see in the next section, our results agree 
with their conclusions, in the sense that the M31 PN display the same correlations as type II PN 
in the Galaxy. These results have been complemented with data on 2 PN by Balick et al. (\citeyear{balick}).  
HII regions in this object have been analyzed by Zurita \& Bresolin (\citeyear{zurita}), Sanders et al. 
(\citeyear{sanders}) and Esteban et al. (\citeyear{esteban2009}), usually by the direct method involving 
the determination of the electron temperature. 

Richer \& McCall (\citeyear{richer2008}) studied a sample of 14 PN in M32, 4 PN in NGC 185 and 10 PN in 
NGC 205. Some enrichment was observed especially in N in the bright PN, suggesting that the main sequence 
mass of the PN progenitors was about 1.5 solar masses or less. 

For M33, Bresolin et al. (\citeyear{bresolin2010}) have obtained accurate abundances for 16 PN and 3 HII 
regions near the central parts of the galaxy. They have found similar trends as those shown in section 3, 
including a comparison with BCG/ELG from Izotov et al. (\citeyear{izotov2006}). For this object, data from 
Magrini et al. (\citeyear{magrini2009}) for a large sample of PN were also considered. Also 
for HII regions, Magrini et al. (\citeyear{magrini2007}) presented data for 14 objects; two more objects 
come from the sample by Esteban et al. (\citeyear{esteban2009}), while Rosolowsky \& Simon  
(\citeyear{rosolowsky}) analyzed a larger sample of 60 HII regions with oxygen abundances. 

For M51, Bresolin et al (\citeyear{bresolin2004}) presented a sample of 10 HII regions. For M81 we have 
used HII region data by Stanghellini et al. (\citeyear{stanghellini}), and for M101 we have used HII region 
data by Kennicutt et al. (\citeyear{kennicutt}) and Esteban et al. (\citeyear{esteban2009}).

\begin{table*}\centering
\small
\caption[]{Data for planetary nebulae.}
\label{table1}
\begin{flushleft}
\begin{tabular}{clcl}
\noalign{\smallskip}
\hline\noalign{\smallskip}
 & System  & number & Reference \\
\noalign{\smallskip}
\hline\noalign{\smallskip}
 1  & MW disk  & 230 & IAG \\
 2  & MW bulge &	179 & IAG \\
 3  & MW disk  &	22  & Chiappini et al. (2009)\\
 4  & MW bulge & 88 & Chiappini et al. (2009) \\
 5  & MW       & 19	 & Girard et al. (2007) \\
 6  & MW       & 4 & Henry et al. (2004) \\
 7  & MW       & 20 & Henry et al. (2010) \\
 8  & MW       &     2 & Milingo et al. (2010) \\
 9  & MW       &     6 & Milanova \& Kholtygin (2009) \\
10  & MW       & 44 & Garc\'\i a-Hern\'andez \& G\'orny (2014)\\
11  & LMC      & 251 & IAG, SRM, LD \\
12  & LMC      & 110 & Chiappini et al. (2009) \\
13  & LMC      & 14 & Milanova \& Kholtygin (2009)\\
14  & SMC      & 129 & IAG, SRM, LD \\
15  & SMC      & 7 & Milanova \& Kholtygin (2009) \\
16  & M31      & 1 & Balick et al. (2013) \\
17  & M31      & 16 & Kwitter et al. (2012)\\
18  & M32      & 14 & Richer \& McCall (2008)\\
19  & M33      & 16 & Bresolin et al. (2010)\\
20  & M33      & 93 & Magrini et al. (2009)\\
21  & NGC 185  & 5 & Richer \& McCall (2008) \\
22  & NGC 205  & 10 & Richer \& McCall (2008) \\
23  & NGC 300  & 25 & Stasi\'nska et al. (2013)\\
24  & NGC 3109 & 7 & Pe\~na et al. (2007) \\
25  & Sextans  & 6 & Magrini et al. (2005)\\
\noalign{\smallskip}
\hline
\end{tabular}
\end{flushleft}
\end{table*}
\begin{table*}\centering
\small
\caption[]{Data for HII regions, BCG, and ELG.}
\label{table2}
\begin{flushleft}
\begin{tabular}{clcl}
\noalign{\smallskip}
\hline\noalign{\smallskip}
 & System  & number & Reference \\
\noalign{\smallskip}
\hline\noalign{\smallskip}
1  & MW       & 34  & Afflerbach et al. (1997)          \\
2  & MW	      & 53  & Guseva  et al. (2007)             \\
3  & MW       & 123 & Rudolph et al. (2006)             \\
4  & MW       & 2   & Tsamis et al. (2003)              \\
5  & MW	      & 4   & Delgado-Inglada et al. (2015)     \\
6  & MC       & 3   & Tsamis et al. (2003)              \\
7  & MC       & 12  & Reyes et al. (2015)               \\
8  & MC       & 1   & Peimbert et al. (2000)            \\
9  & MC       & 2   & Pe\~na-Guerrero et al. (2012)     \\
10 & MC       & 2   & Selier \& Heydari-Malayeri (2012) \\
11 & MC       & 15  & Vermeij \& van der Hulst (2002)             \\
12 & M31      & 1   & Esteban et al. (2009)             \\
13 & M31      & 52  & Sanders et al. (2012)             \\
14 & M31      & 9   & Zurita \& Bresolin (2012)         \\
15 & M33      & 3   & Bresolin et al. (2010)            \\
16 & M33      & 2   & Esteban et al. (2009)             \\
17 & M33      & 14  & Magrini et al. (2007)             \\
18 & M33      & 60  & Rosolowsky \& Simon (2008)        \\
19 & M51      & 10  & Bresolin et al. (2004)            \\ 
20 & M81      & 12  & Stanghellini et al. (2014)        \\
21 & M101     & 4   & Esteban et al. (2009)             \\
22 & M101     & 20  & Kennicutt et al. (2003)           \\
23 & NGC 300  & 28  & Bresolin et al. (2009)            \\
24 & NGC 300  & 9   & Stasi\'nska et al. (2013)         \\
25 & NGC 628  & 45  & Berg et al. (2015)                \\
26 & NGC 3109 & 10  & Pe\~na et al. (2007)              \\
27 & NGC 5194 & 29  & Croxall et al. (2015)             \\
28 & Sextans  & 17  & Magrini et al. (2005)             \\
29 & BCG      & 54  & Izotov \& Thuan (1999)            \\
30 & ELG      & 306 & Izotov et al. (2006)              \\
\noalign{\smallskip}
\hline
\end{tabular}
\end{flushleft}
\end{table*}
\begin{table*}
\small
\caption[]{Total samples.}
\label{table3}
\begin{flushleft}
\begin{tabular}{lrlr}
\noalign{\smallskip}
\hline\noalign{\smallskip}
PLANETARY NEBULAE       & Number & \ HII REGIONS             & Number \\
\hline\noalign{\smallskip}
Milky Way Disk          & 347    & \ Milky Way               & 216 \\
Milky Way Bulge         & 267    & \ Magellanic Clouds       & 35  \\
Milky Way               & 614    & \ Other Galaxies          & 325 \\
Magellanic Clouds       & 511    & \ BCG, ELG                & 360 \\
Total External Galaxies	& 704    & \ Total External Galaxies & 720 \\
TOTAL                   & 1318   & \ TOTAL                   & 936 \\
\noalign{\smallskip}
\hline
\end{tabular}
\end{flushleft}
\end{table*}

For NGC 300, Stasi\'nska et al. (\citeyear{stasinska2013}) have analyzed a sample of PN and compact HII 
regions, and a similar sample of HII regions was taken from Bresolin et al. (\citeyear{bresolin2009}) 
for this galaxy. For NGC 628 we have used recent data for a large sample of HII regions by Berg et al. 
(\citeyear{berg}).

In the case of NGC 3109, Pe\~na et al. (\citeyear{pena2007}) studied a sample of PN and HII regions and 
concluded that O and Ne abundances may be affected by the evolution of the PN central stars, although 
the size of the sample and the intrinsic uncertainties make this conclusion controversial. For NGC 5194, 
recent results on HII regions by Croxall et al. (\citeyear{croxall}) have been taken into account. 
Finally, for the Sextans A and B galaxies, we have used data by Magrini et al. (\citeyear{magrini2005}), 
who have  analyzed samples of PN and HII regions in these objects.

We have also included in our analysis a sample of Blue Compact Galaxies and Emission Line Galaxies 
from Izotov and Thuan (\citeyear{izotov1999}) and Izotov et al. (\citeyear{izotov2006}), as they are 
equivalent of HII regions at relatively low metallicities (cf. Bresolin et al. \citeyear{bresolin2010}),
which will be confirmed in the next section. These objects are essentially giant metal-poor HII regions, 
and fit nicely in the correlations obtained, but concentrate towards lower metallicities, $\log{\rm (O/H)}
+ 12 \leq  8.2$.

Tables 1 and 2 summarize the data for PN and HII regions, respectively, and Table 3 shows the total number
of objects in each class. 

\section{Results}
\label{section3}

\subsection{Elements not  produced by the PN  progenitor stars}
\label{subsection31}

The abundances of the elements O, Ne, S, and Ar are probably not significantly affected by the evolution 
of the PN progenitor stars, and the corresponding distance-independent correlations are well determined. The 
measured abundances therefore reflect the interstellar abundances at the time the progenitor stars were 
born and can be compared with the data for the younger objects, such as HII regions. Oxygen can be used 
as a metallicity proxy, and  an  accurate relation between O and Fe can be determined (see for example 
Ram\'\i rez et al. \citeyear{ramirez}). This is important, since in photoionized nebulae Fe is mostly locked 
up in grains. The average slope of the Fe - O correlation is approximately of 1.11 for the thin disk and 
1.31 for the thick disk. 

\begin{figure}
   \centering
   \includegraphics[angle=-90, width=5.5cm]{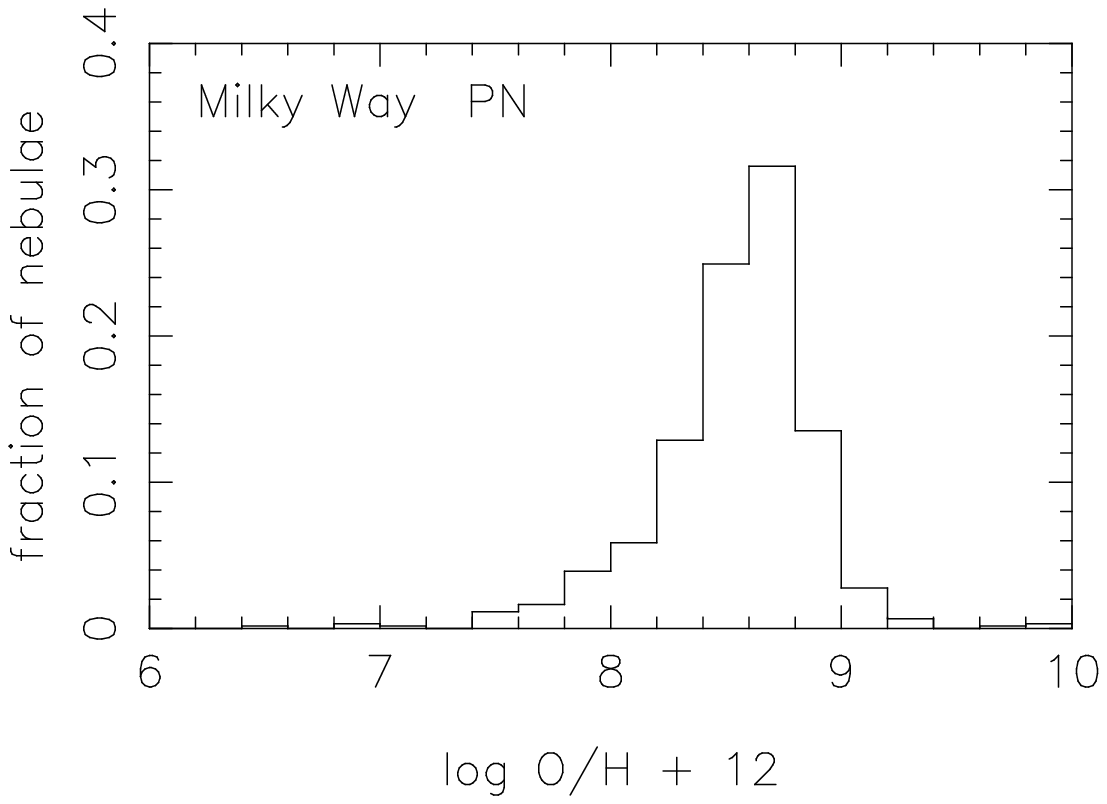}
   \includegraphics[angle=-90, width=5.5cm]{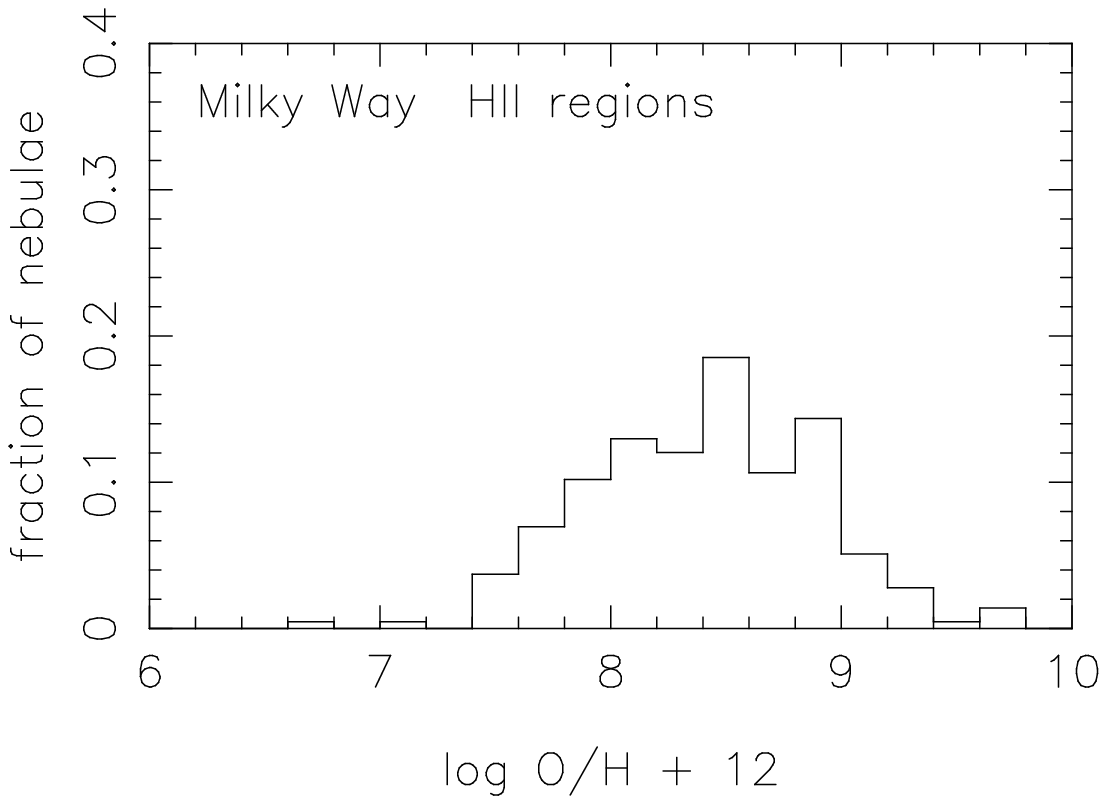}
   \includegraphics[angle=-90, width=5.5cm]{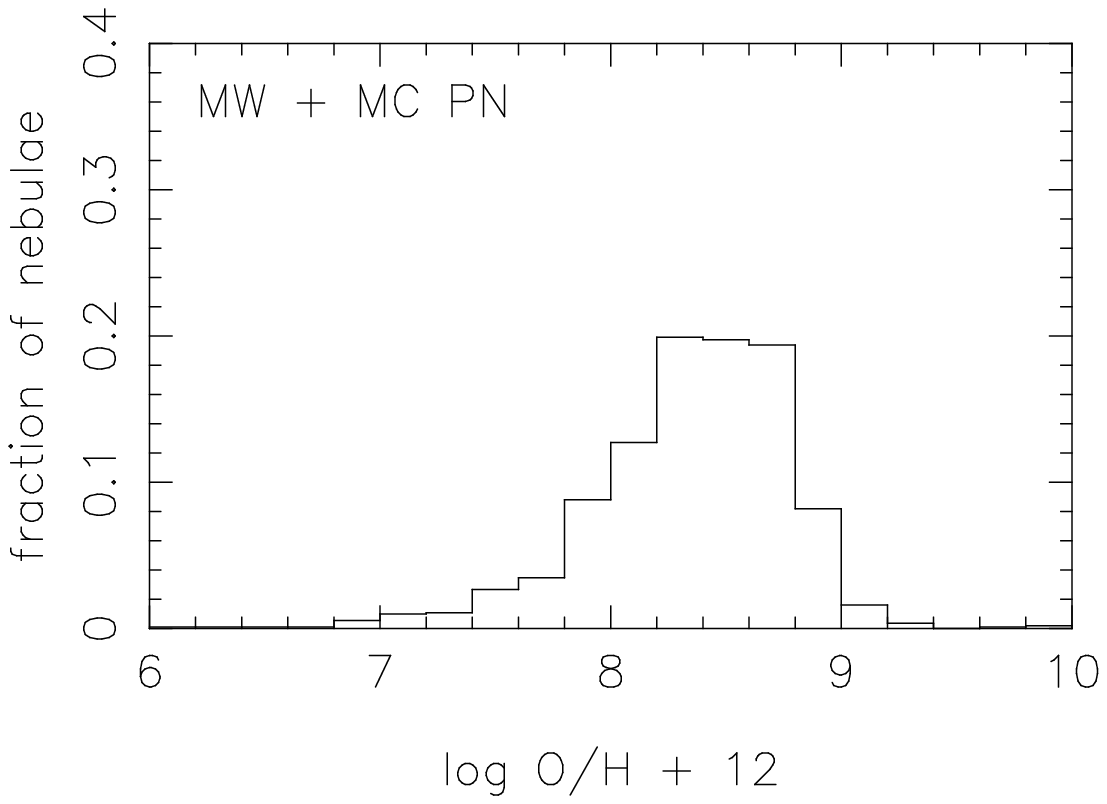}
   \includegraphics[angle=-90, width=5.5cm]{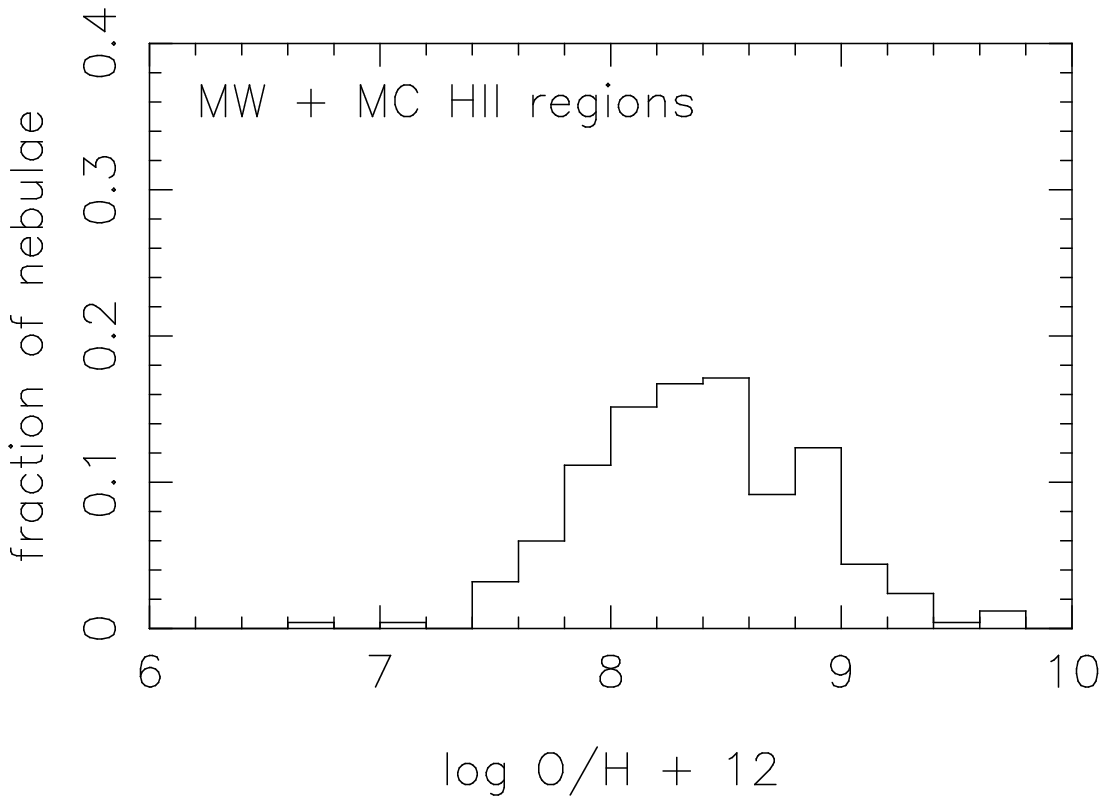}
   \includegraphics[angle=-90, width=5.5cm]{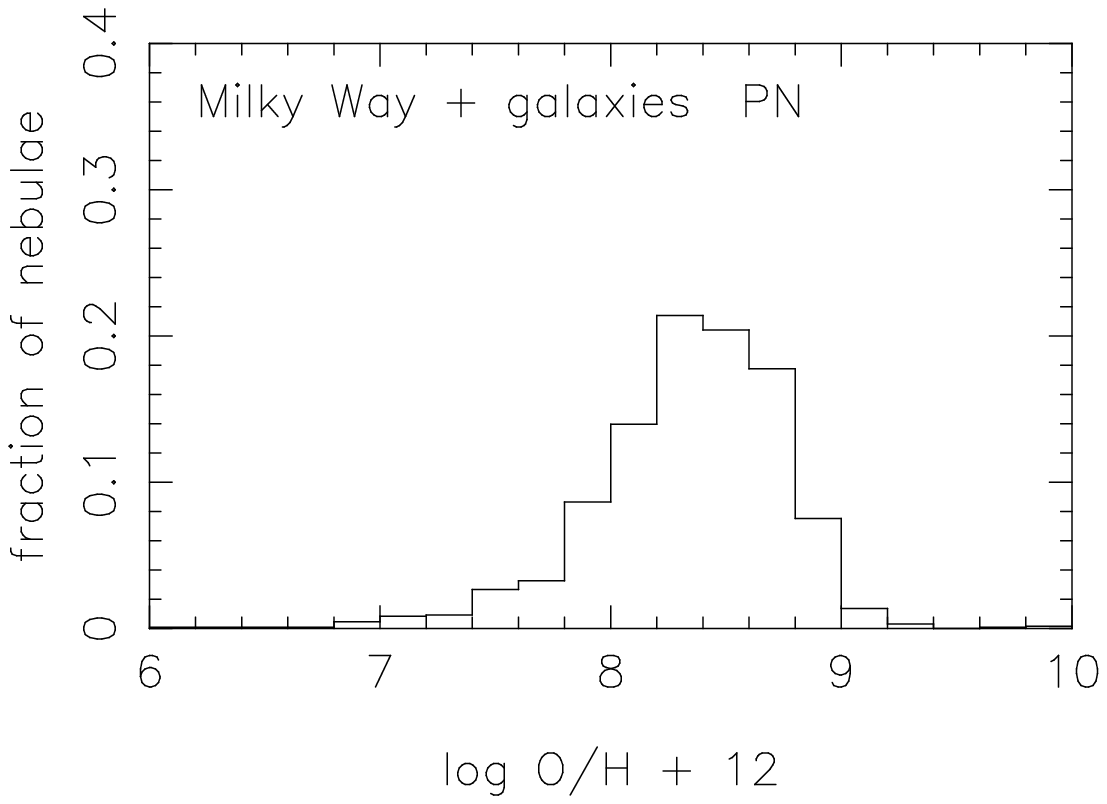}
   \includegraphics[angle=-90, width=5.5cm]{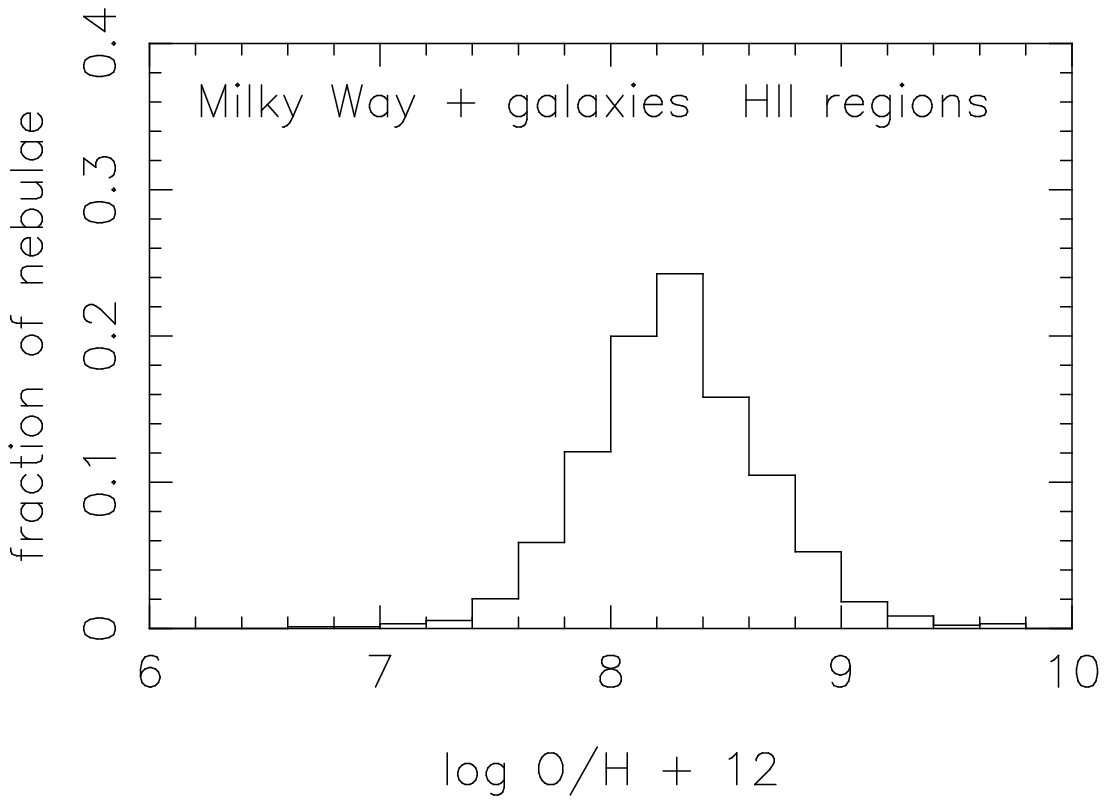}
   \caption{Histograms of the O/H abundances in PN and HII regions. top: Milky Way;
   middle: Milky Way and Magellanic Clouds; bottom: Milky Way and all external galaxies.}
   \label{fig1}
   \end{figure}

Figure \ref{fig1} shows histograms of the oxygen abundance O/H for planetary nebulae and HII regions in
three representative cases: The Mllky Way (top figures), the Milky Way and the Magellanic Clouds
(middle figures) and all objects considered here, namely the Milky Way, the Magellanic Clouds
and the remaining external galaxies (bottom figures). It can be seen that both PN and HII regions
have similar distributions, peaking around $\log {\rm (O/H)} + 12 = 8.4 \ {\rm to}\  8.8$, although
the HII region distributions are generally broader than in the case of planetary nebulae.
Also, it can be noted that a larger fraction of HII regions have $\log {\rm (O/H)} + 12 \geq 9$,
which reflects the fact that these younger objects are formed by more enriched material. 
Similar plots can be obtained for Ne, S, and Ar, although the samples are smaller than in the case
of oxygen, since the abundances of these elements are usually more difficult to obtain. For the S/H ratio 
similar peaks are also observed for PN and HII regions in the three cases considered, but for Ne/H and 
Ar/H the total number of HII regions is relatively small, so that a comparison is difficult. An exception 
is the Ar/H histograms for all objects, which is similar to the results of Figure~\ref{fig1} (bottom).

In this paper we concentrate on distance-independent correlations, to avoid  distance determination 
problems, which affect processes such as the abundance gradients observed in the Milky Way 
and other spirals. The main results are shown in Figures 2 to 10 for neon, sulphur, and argon,   
as functions of the oxygen abundance by number of atomos, O/H.  The  top figures show the abundances 
of these elements relative to hydrogen (Ne/H, S/H, and Ar/H), while the bottom figures show the 
abundances relative to oxygen (Ne/O, S/O, and Ar/O). In these figures, PN are always represented by 
empty symbols (squares and circles), while HII regions, BCG, and ELG are represented  by triangles or crosses.

\bigskip\noindent
NEON
\bigskip

\begin{figure}
   \centering
   \includegraphics[angle=0, width=13.0cm]{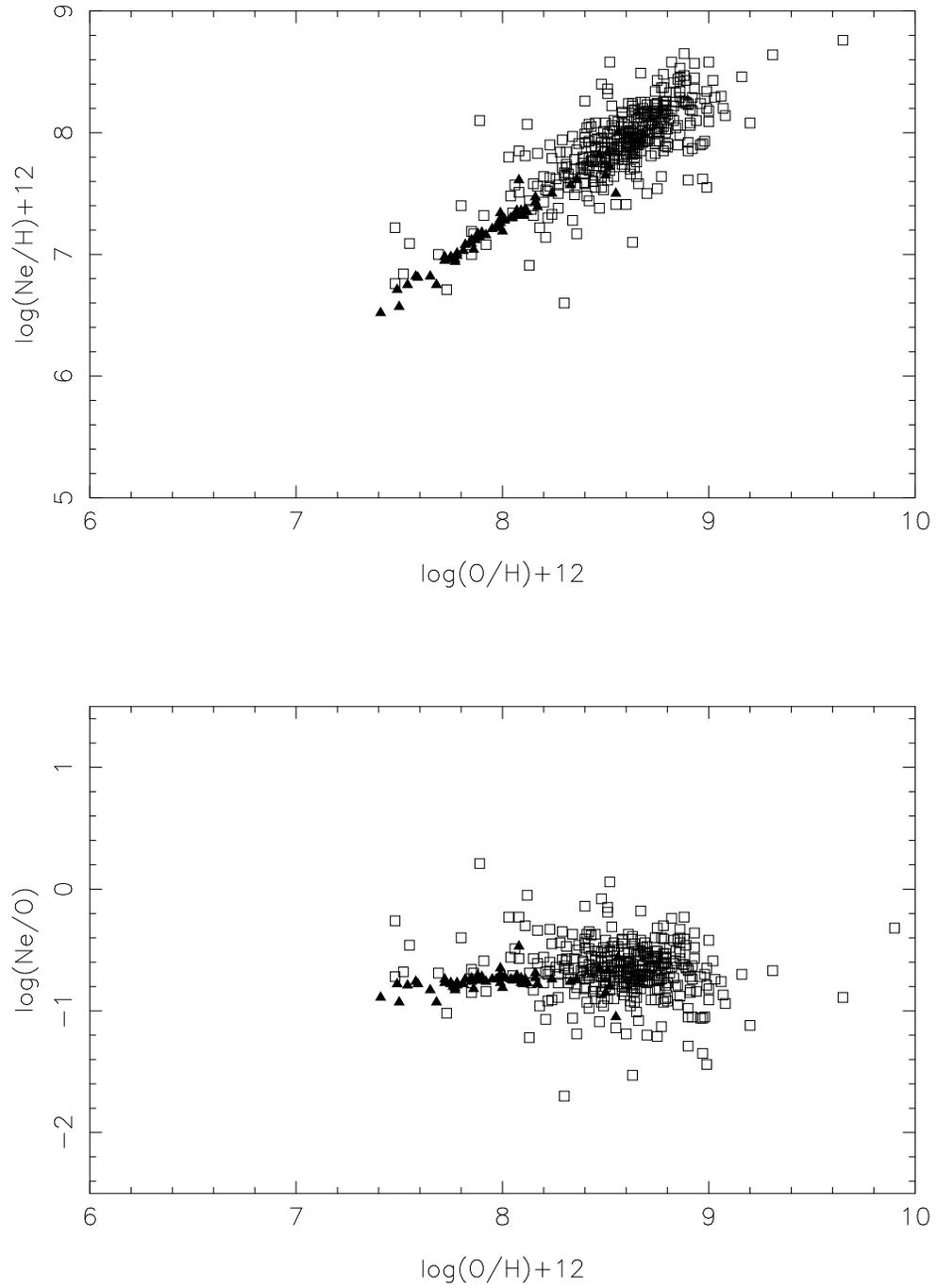}
   \caption{Ne abundances as functions of oxygen abundances for the Milky Way: PN (squares),
    HII regions (triangles).}
   \label{fig2}
   \end{figure}

\begin{figure}
   \centering
   \includegraphics[angle=0, width=13.0cm]{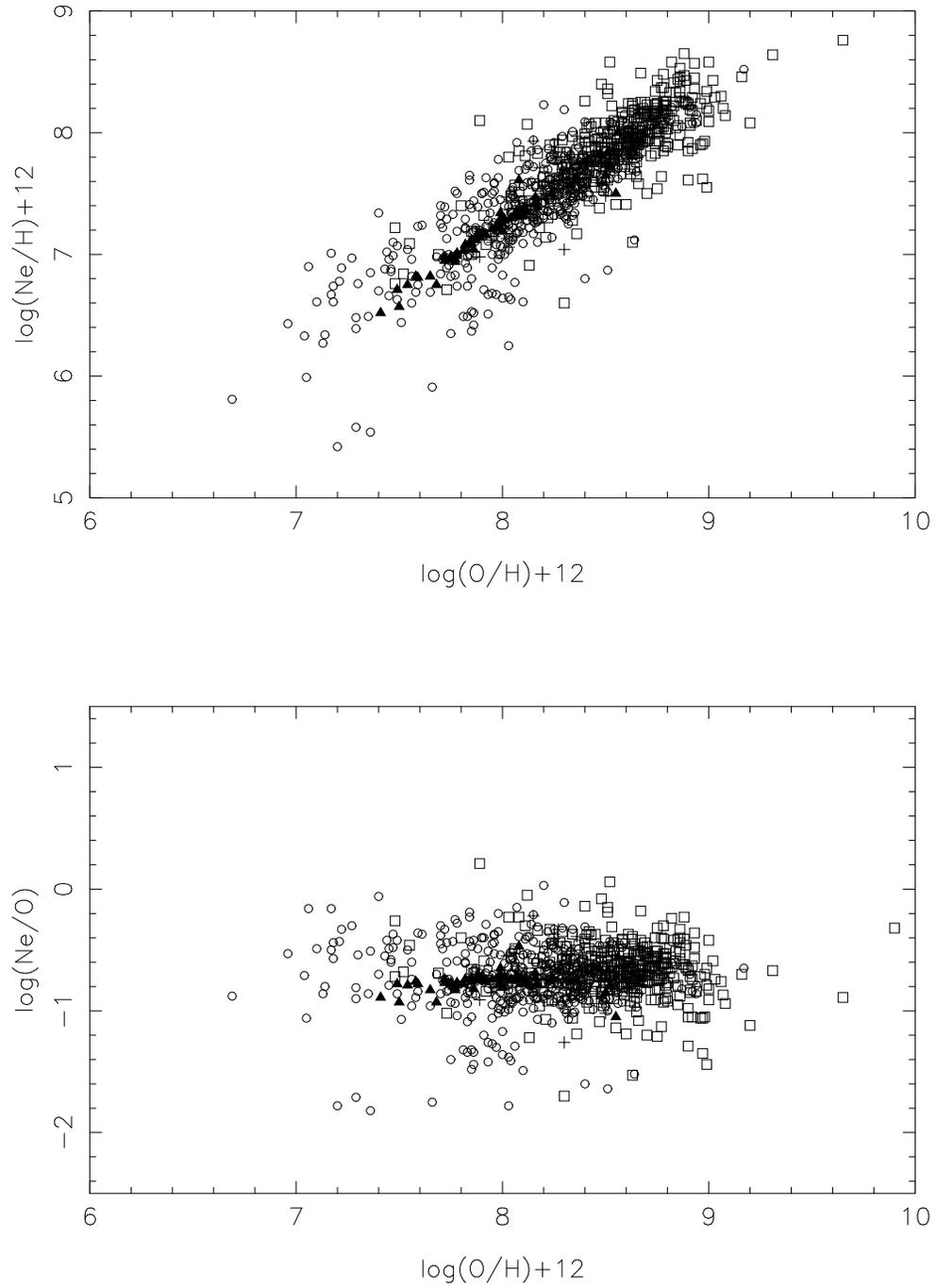}
   \caption{The same as Figure \ref{fig2} for the Milky Way and Magelllanic 
   Clouds:  MW PN (squares), MW HII regions (triangles), MC PN (circles), MC HII regions (crosses).}
   \label{fig3}
   \end{figure}

\begin{figure}
   \centering
   \includegraphics[angle=0, width=13.0cm]{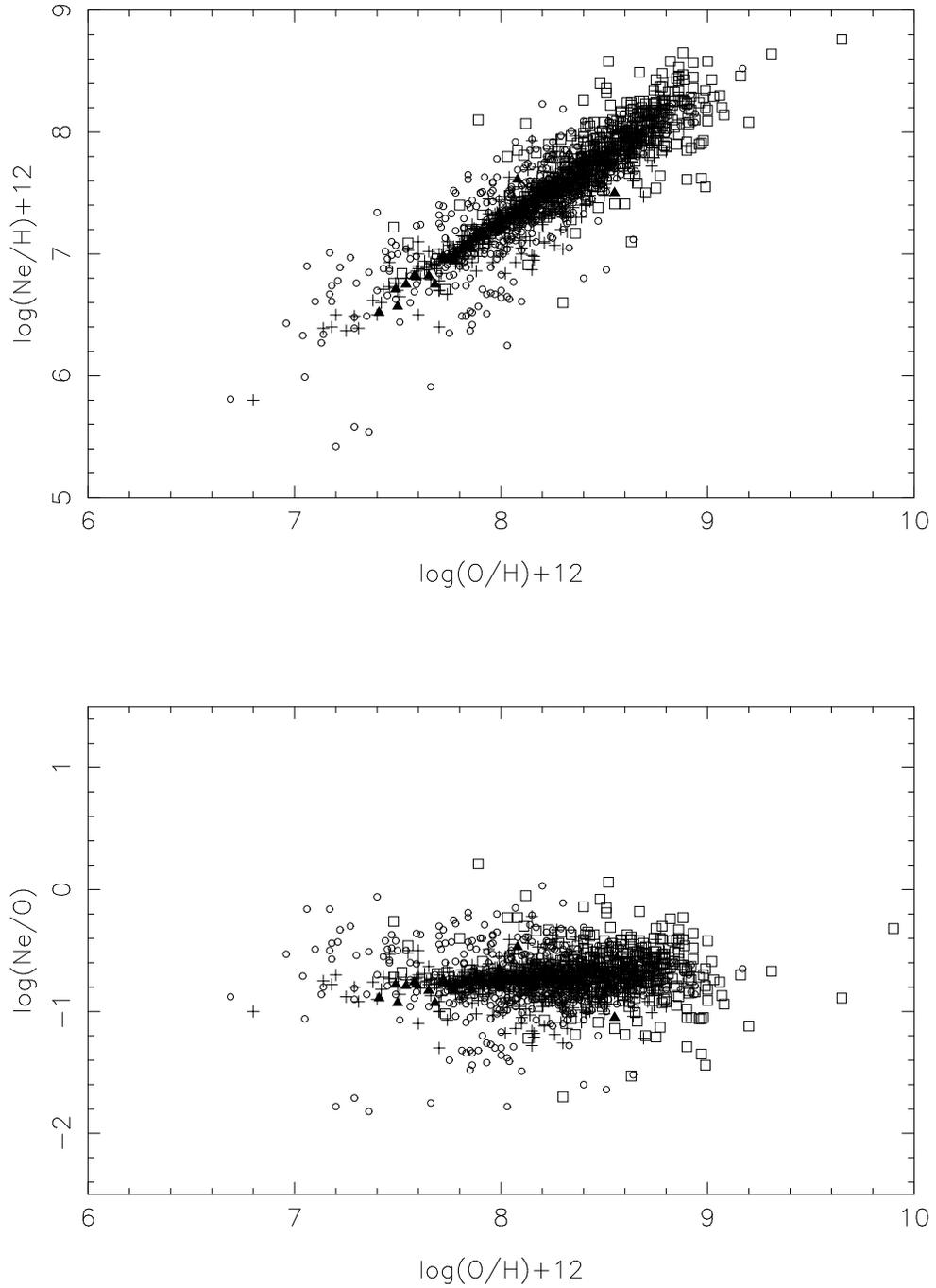}
   \caption{The same as Figure \ref{fig2} for the Milky Way and external 
    galaxies: MW PN (squares), MW HII regions (triangles), external PN (circles), external 
    HII regions (crosses).}
   \label{fig4}
   \end{figure}

The results for Ne are shown in Figures \ref{fig2}, \ref{fig3}, and \ref{fig4}. Figure \ref{fig2} shows the 
Milky Way objects, where the squares represent PN and the triangles represent HII regions. Bulge PN and disk 
PN have a similar behavior, so that they are both included as squares in the figure. From the top figure we 
see that for the Ne/H ratio both PN and HII regions present a lockstep variation with O/H, although for HII 
regions the dispersion is much smaller than in the case of planetary nebulae. This is also reflected in the 
bottom figure, which shows the Ne/O ratio as a function of O/H, indicating that the Ne/O ratio is essentially 
constant with a higher dispersion for planetary nebulae. The estimated dispersions are given in Table~4
for the correlations of the abundances relative to hydrogen as functions of the O/H ratio.

%
\begin{table}
\small
\caption[]{Average dispersions as functions of O/H.}
\label{table4}
\begin{flushleft}
\begin{tabular}{lccc}
\noalign{\smallskip}
\hline\noalign{\smallskip}
 & MW & MW+MC & ALL \\
\hline\noalign{\smallskip}
N/H         &      &      &      \\			
PN          & 0.42 & 0.48 & 0.38 \\
HII regions & 0.33 & 0.41 & 0.31 \\
ALL         & 0.42 & 0.49 & 0.41 \\
            &      &      &      \\
Ne/H        &      &      &      \\			
PN          & 0.22 & 0.25 & 0.20 \\
HII regions & 0.09 & 0.15 & 0.13 \\
ALL         & 0.21 & 0.24 & 0.16 \\
            &      &      &      \\			
S/H         &      &      &      \\			
PN          & 0.33 & 0.38 & 0.27 \\
HII regions & 0.18 & 0.17 & 0.15 \\
ALL         & 0.30 & 0.36 & 0.20 \\
            &      &      &      \\			
Ar/H        &      &      &      \\			
PN          & 0.34 & 0.33 & 0.33 \\ 
HII regions &      & 0.11 & 0.17 \\
ALL         & 0.34 & 0.32 & 0.24 \\
\noalign{\smallskip}
\hline
\end{tabular}
\end{flushleft}
\end{table}

Data for the Magellanic Clouds are included in Figure \ref{fig3}, where the squares are again PN in the 
Milky Way, the circles are PN in the Magellanic Clouds, the triangles represent HII regions in the 
Milky Way and the crosses are HII regions in the Magellanic Clouds. There is a well known metallicity 
difference between the LMC and SMC in the sense that the LMC has a higher average metallicity than the 
SMC, which is not shown in the figure, as both Clouds are represented by the same symbols. We can see 
that the same trend of Figure \ref{fig2} is also apparent here, in the sense that the Ne abundances vary
in lockstep with oxygen. A similar dispersion is observed for PN in both Figures \ref{fig2} and
\ref{fig3}.  The MC data show a rather smooth transition to 
the Galaxy towards lower metallicities, which is especially noticeable in the PN data, where the separation 
of the MW PN and MC PN is more evident. The Magellanic Cloud PN are displaced towards lower Ne abundances, 
which reflects the lower metallicity of the Clouds relative to the Milky Way.
 
Data for the remaining external galaxies, comprising all galaxies in the Local Group, BCG, and ELG, are 
included in Figure \ref{fig4}. Here again MW PN are shown as squares, PN in external galaxies are shown as 
circles, MW HII regions are represented by triangles, and external HII regions are shown as crosses.  It is 
remarkable that the same behaviour observed in the Galaxy also holds in other Local Group objects. Despite 
their different metallicities and morphologies, their nucleosynthetic processes and chemical evolution are 
apparently very similar. The trend displayed in Figure \ref{fig4} (top) shows a very good agreement with the 
trend found by Izotov et al. (\citeyear{izotov2006}) on the basis of ELG only. Similar conclusions were obtained 
by Richer \& McCall (\citeyear{richer2007}, \citeyear{richer2008}). Even the average dispersions do not change 
appreciably for Local Group objects, and the lower dispersion in HII region abundances compared with PN has the 
same characteristics both in the Milky Way and in the Local Group objects. The fractions of objects within 
1$\sigma$, 2$\sigma$, and 3$\sigma$ are shown in Table~5, where both PN and HII regions are included.

\begin{table}
\small
\caption[]{Fractions of objects within 1$\sigma$, 2$\sigma$, and 3$\sigma$.}
\label{table5}
\begin{flushleft}
\begin{tabular}{lccc}
\noalign{\smallskip}
\hline\noalign{\smallskip}
 & 1$\sigma$ & 2$\sigma$ & 3$\sigma$ \\
\hline\noalign{\smallskip}
Ne/H          &      &      &      \\			
Milky Way     & 0.75 & 0.94 & 1.00 \\
MW + MC       & 0.77 & 0.94 & 1.00 \\
MW + galaxies & 0.78 & 0.93 & 1.00 \\
              &      &      &      \\			
S/H           &      &      &      \\			
Milky Way     & 0.36 & 0.58 & 1.00 \\
MW + MC       & 0.54 & 0.84 & 1.00 \\
MW + galaxies & 0.30 & 0.56 & 1.00 \\
              &      &      &      \\			
Ar/H          &      &      &      \\			
Milky Way     & 0.41 & 0.67 & 1.00 \\ 
MW + MC       & 0.54 & 0.85 & 1.00 \\
MW + galaxies & 0.44 & 0.73 & 1.00 \\
\noalign{\smallskip}
\hline
\end{tabular}
\end{flushleft}
\end{table}

It is also interesting to notice that in this larger sample the observed ranges of oxygen and neon abundances 
are similar, in spite of a few PN at very high oxygen abundances. The similarity essentially reflects the 
fact that the interstellar metallicities did not change appreciably in the last 5 Gyr approximately, a result 
that is supported by recent determinations of the age-metallicity in the Milky Way (see for example Rocha-Pinto
et al. \citeyear{rocha-pinto}, Bensby et al. \citeyear{bensby2004a}, \citeyear{bensby2004b}). The abundances 
relative to oxygen shown in the bottom figure remain  essentially constant, as in the case of Figures 
\ref{fig2} and \ref{fig3}, but the dispersion of PN data is considerably higher. 

The  similarity of all photoionized nebulae coupled with the high dispersions shown by PN can be 
interpreted assuming that the dispersion in the PN data reflects the fact that the abundances are not 
as well determined as in the HII regions. However, a larger dispersion would be expected, since PN are 
older objects than the HII regions and any given sample probably includes objects of different ages, as 
we have shown elsewhere (Maciel et al. \citeyear{maciel2010a}, \citeyear{maciel2010b}, 
\citeyear{maciel2011}). Apart from a few objects at very low metallicities, the same dispersion is 
observed at all metallicities, amounting to about 0.3 dex, considering also S and Ar. This is higher than 
or similar to the individual uncertainties in the abundances, so that the dispersion is probably real, 
reflecting the different ages of the PN central stars.

Alternatively, it can be considered that there is some contribution to the Ne abundances from the PN 
progenitor stars, as suggested in some recent investigations (see for example Pe\~na et al. 
\citeyear{pena2007}). Although oxygen and neon are mostly produced by the evolution of massive stars through 
core-collapse supernovae, some contribution from the PN progenitor stars is also supported by theoretical models.
This interpretation, however, conflicts with the fact that the dispersion does 
not seem to change for different metallicities. If an important part of the Ne abundances were produced 
by the intermediate stars we would expect the dispersion to increase with the metallicity, which is not 
observed. Therefore, the most accurate data available for Ne do not suggest any important contribution 
from the PN progenitor stars, so that any such contribution would be expected to be smaller than the average 
uncertainties in the abundances. 

It has been argued that  the third dredge-up process in AGB stars may affect the oxygen abundances 
observed in planetary nebulae (see for example Karakas \& Lattanzio \citeyear{karakas2014}).  ON cycling 
would also reduce the O/H ratio especially in lower metallicity PN. This appears to affect low metallicity 
objects with progenitor star masses higher than about $2 M_\odot$ (see for example Karakas \& Lattanzio 
\citeyear{karakas2007}, Herwig  \citeyear{herwig}). Our results show that, if present, such contribution 
should be small compared with the average uncertainties in the PN abundances.
Figures \ref{fig2} to \ref{fig4} show some evidence of this effect for $\log {\rm (O/H)} + 12 \leq 8$, but 
the number of PN at such low metallicities is a small fraction of the sample considered. It can be concluded 
that this effect, if present, is presently masked by the uncertainties in the abundance determinations, as 
well as by the different average metallicities of the objects considered here.

Taking into account the expected  uncertainties in these abundances, an average contribution 
of about 0.1 dex cannot be ruled out. On the other hand, if we compare the expected contributions both to 
oxygen and neon, it is unlikely that their are equal, which is needed in order to explain the similarity of 
the PN and HII region trends shown in Figures \ref{fig2} to \ref{fig4} (cf. Karakas and Lattanzio 
\citeyear{karakas2003}).

\bigskip\noindent
SULPHUR
\bigskip

\begin{figure}
   \centering
   \includegraphics[angle=0, width=13.0cm]{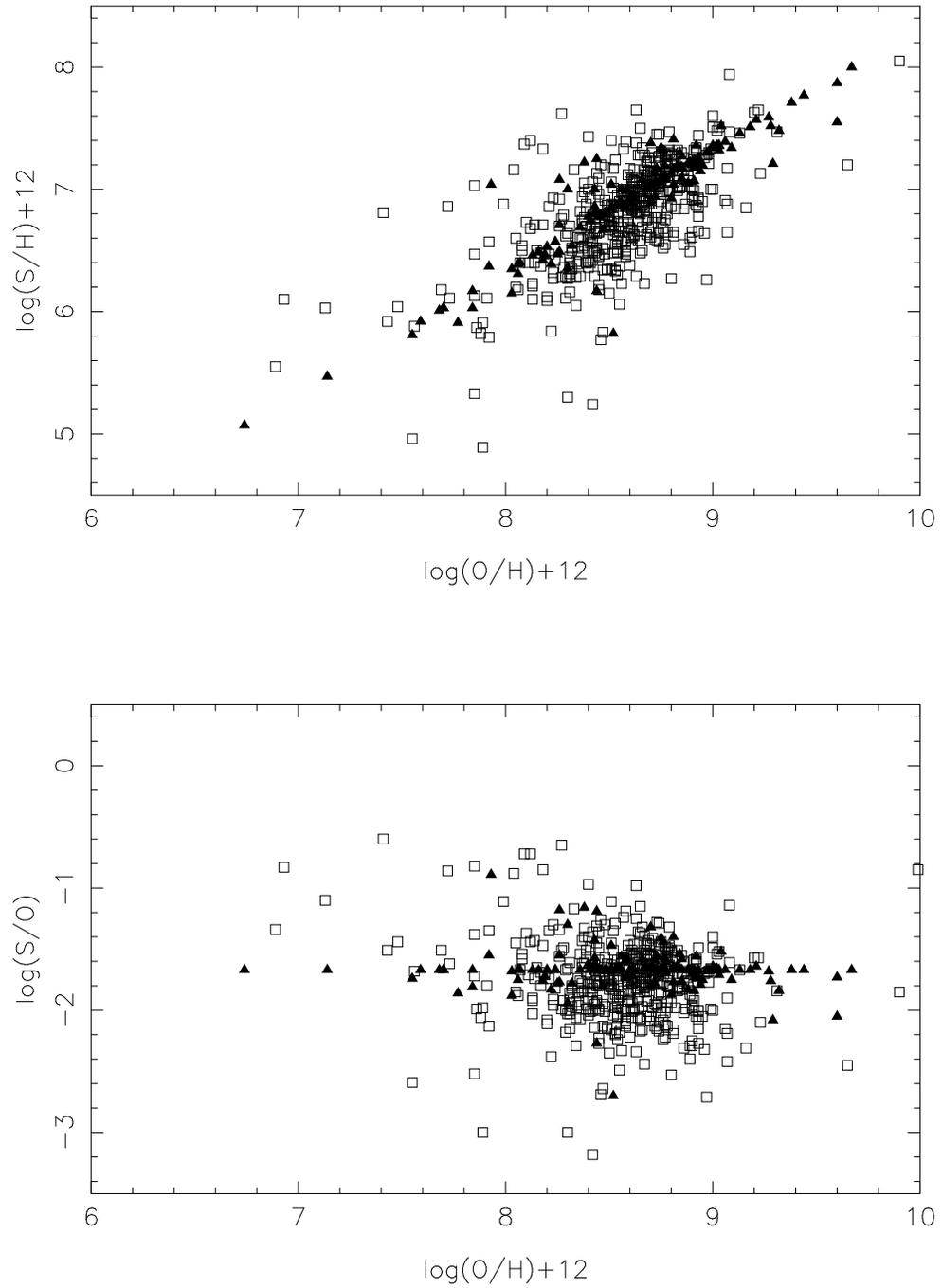}
   \caption{The same as Figure \ref{fig2} for sulphur.}
   \label{fig5}
   \end{figure}

\begin{figure}
   \centering
   \includegraphics[angle=0, width=13.0cm]{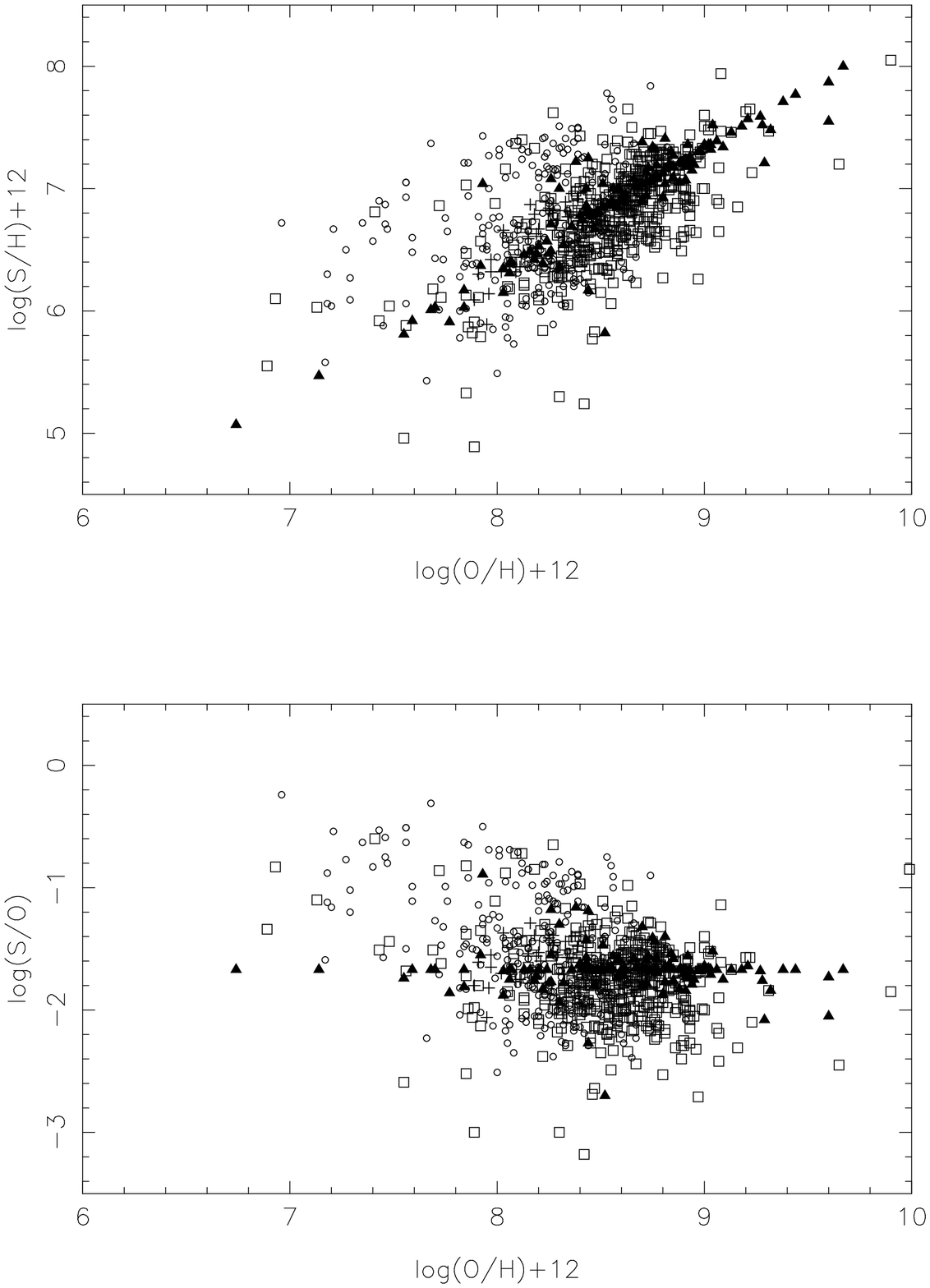}
   \caption{The same as Figure \ref{fig3} for sulphur.}
   \label{fig6}
   \end{figure}

\begin{figure}
   \centering
   \includegraphics[angle=0, width=13.0cm]{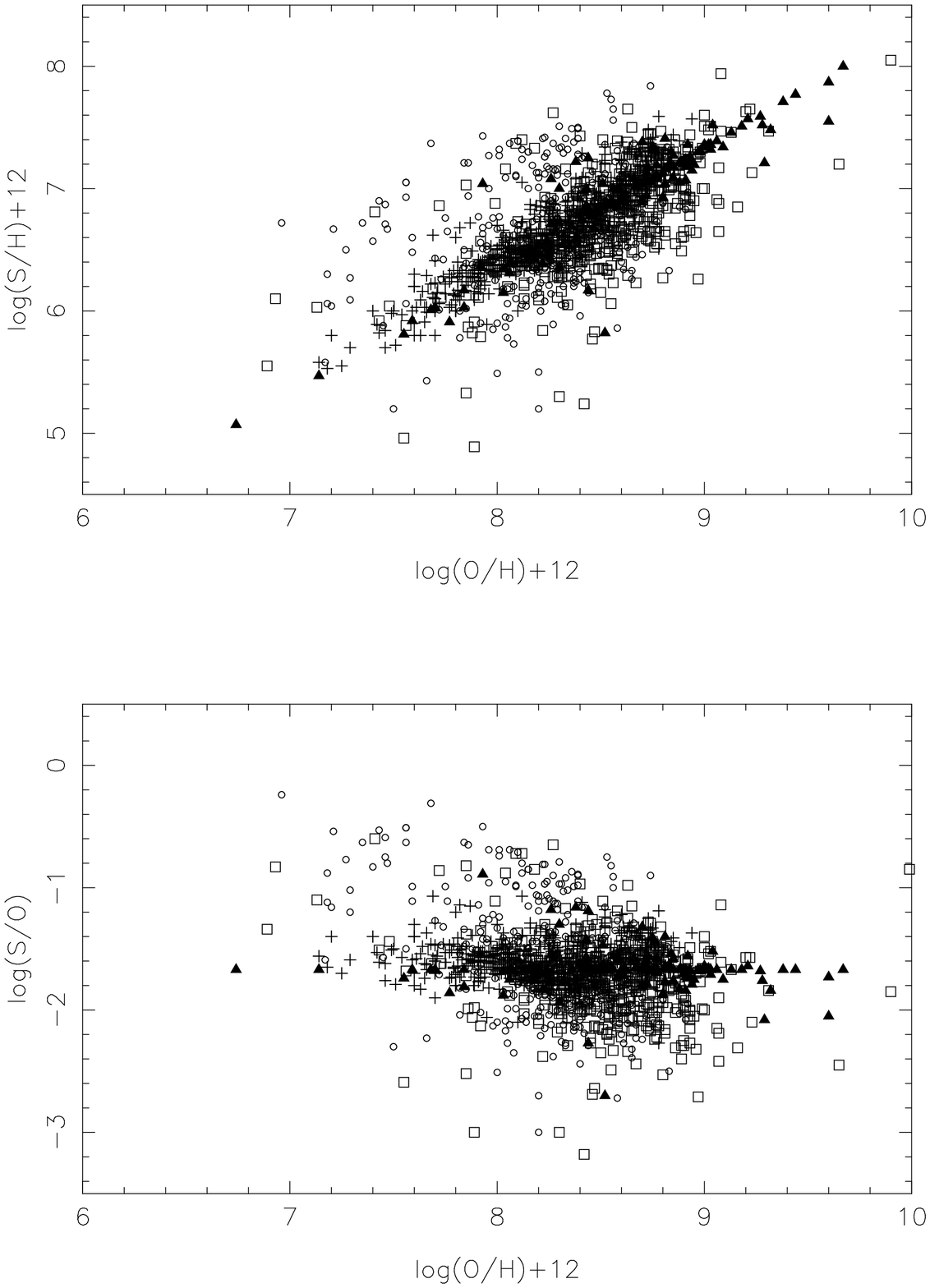}
   \caption{The same as Figure \ref{fig4} for sulphur.}
   \label{fig7}
   \end{figure}

In the case of sulphur, as shown in Figures \ref{fig5}, \ref{fig6}, and \ref{fig7}, the general trends with 
oxygen are similar to neon, but some differences arise. The symbols are as in Figures \ref{fig2}, \ref{fig3}, 
and \ref{fig4}. For the Milky Way disk, the HII regions present a very good correlation, and the data extend 
to higher and lower metallicities compared with neon.  The galactic PN already display what is usually 
called the \lq\lq sulphur anomaly\rq\rq, that is, many PN apparently have somewhat lower S/H abundances than
expected for their metallicity (see for example the detailed discussions by Henry et al. \citeyear{henry2004}, 
\citeyear{henry2012}), which can be observed in Figure \ref{fig5} (top and bottom). The sulphur anomaly has 
been attributed to a deficiency in the calculation of the sulphur ICFs, particularly due to the abundance of 
the S$^{+3}$ ion, lack of accurate atomic constants, effect of the nucleosynthesis in the progenitor stars, 
and different chemical evolution of the systems considered.  In Figure \ref{fig5} it is clearly seen that more 
objects lie below the line defined by the HII regions, especially in the range $8.0 < \log {\rm (O/H)} + 12 < 9.0.$ 

Considering now the Milky Way and the Magellanic Clouds, we see in Figure \ref{fig6}, in which the symbols 
are as in Figure \ref{fig3},  that a new effect is apparent, in the sense that the S abundances in the Magellanic 
Clouds present a much wider dispersion than the galactic objects,  showing many PN with higher sulphur 
abundances than the HII regions. There is still a reasonable number of objects below the HII region curve, 
but there is a large number of PN in the opposite side, so that the sulphur anomaly is not particularly 
noticeable, and in fact the data could be attributed simply to the higher uncertainty for the MC abundances,  
suggesting that these abundances are not as well known as in the galactic PN. 

Taking into account all objects in the Local Group, as shown in Figure~\ref{fig7}, where the symbols are as in 
Figure \ref{fig4}, we see that the same general trends are maintained with a higher dispersion compared with Ne
for PN. It is noteworthy that for all HII regions, galactic as well as extragalactic, the lockstep variation 
with oxygen is very well defined, despite the fact that the objects considered are very different. Another 
consequence is that the sulphur anomaly becomes less important as additional samples are considered. This 
suggests that this problem is very probably related to the calculation techniques used to derive the sulphur 
abundances in galactic and possibly in the Magellanic Cloud PN. 

It can also be seen from Table~4 and Figures \ref{fig5} to \ref{fig7} that the average dispersion of the 
sulphur abundances is higher compared to Ne, and similar to the case of Ar, as we will see later on. 
For HII regions, essentially the same average dispersion is obtained in all cases. The inclusion of BCG and ELG 
maintains these conclusions, that is, the sulphur abundances of HII apparently do not show the sulphur anomaly, 
which is then a characteristic of the empirical determination of sulphur abundances in planetary nebulae. In 
other words, the problem of the sulphur determinations affects basically the planetary nebulae, but not the HII 
regions and blue compact galaxies. 

As a conclusion, we see that the HII region trend is always well defined, while the PN data present at least 
three different problems: (i) the sulphur anomaly, apparent in galactic objects, (ii) a large number of 
objects with higher  sulphur abundances compared with HII regions, especially in the Magellanic Clouds, and 
(iii) a higher dispersion in the PN data, compared to neon. Based on the estimates of the discrepancy 
in the sulphur abundances in  the Clouds and in the Milky Way we suggest that the sulphur abundances of PN 
in the Clouds are probably overestimated up to a factor of the order of 3. Of course, the overestimate may 
be different for individual nebulae, so that this applies only to the bulk of the nebulae. 

\bigskip\noindent
ARGON
\bigskip

\begin{figure}
   \centering
   \includegraphics[angle=0, width=13.0cm]{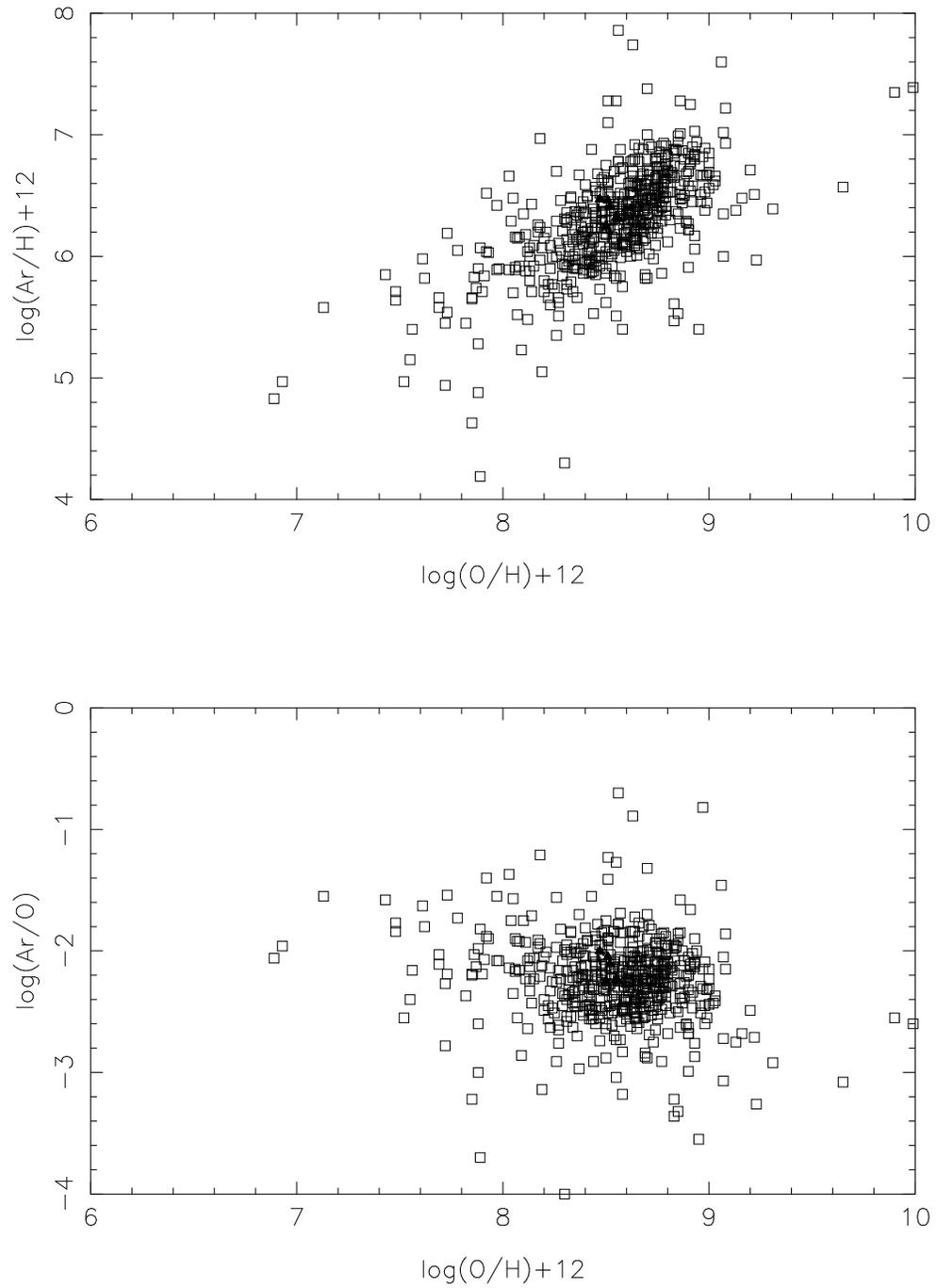}
   \caption{The same as Figure \ref{fig2} for argon.}
   \label{fig8}
   \end{figure}

\begin{figure}
   \centering
   \includegraphics[angle=0, width=13.0cm]{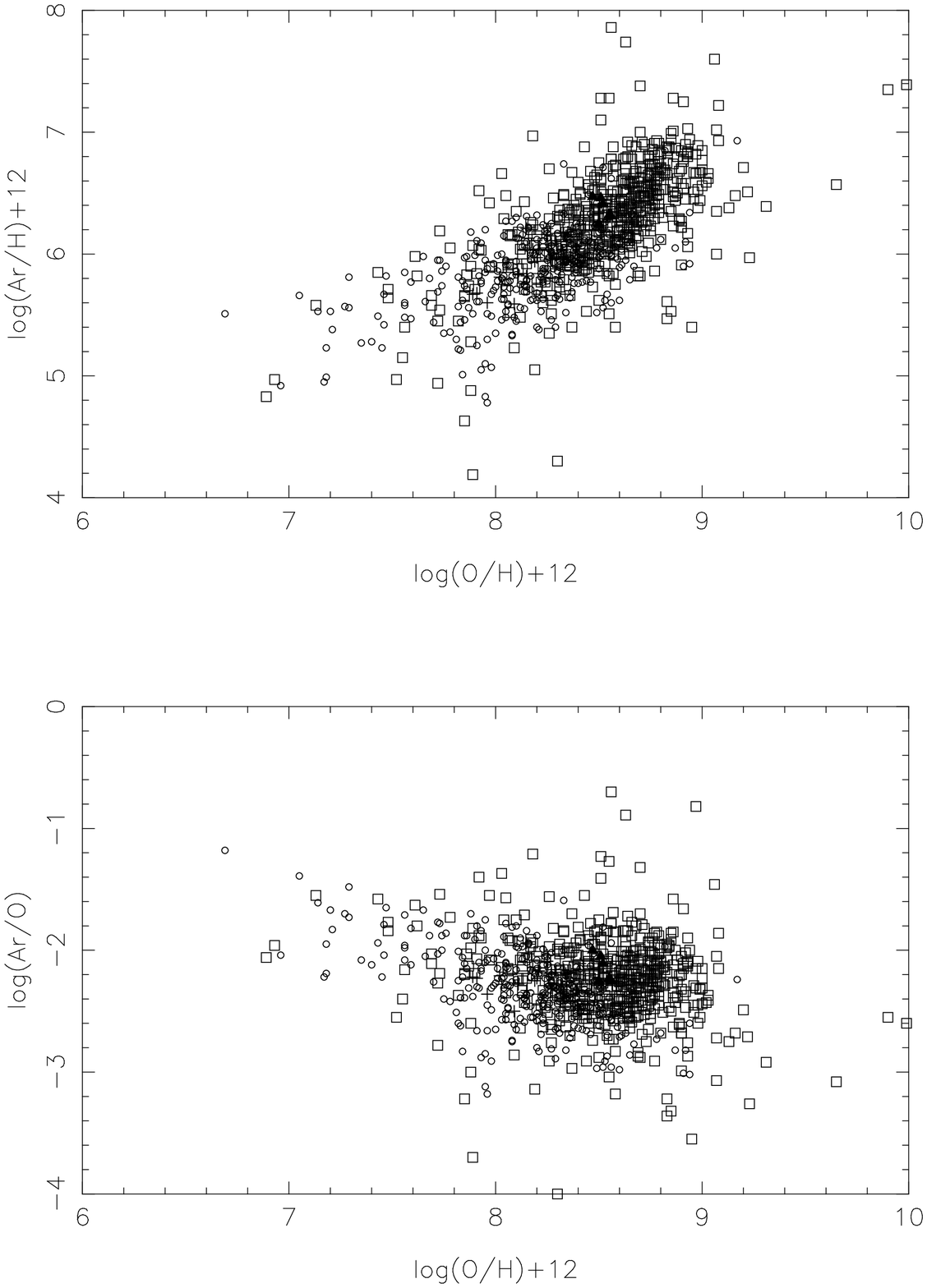}
   \caption{The same as Figure \ref{fig3} for argon.}
   \label{fig9}
   \end{figure}

\begin{figure}
   \centering
   \includegraphics[angle=0, width=13.0cm]{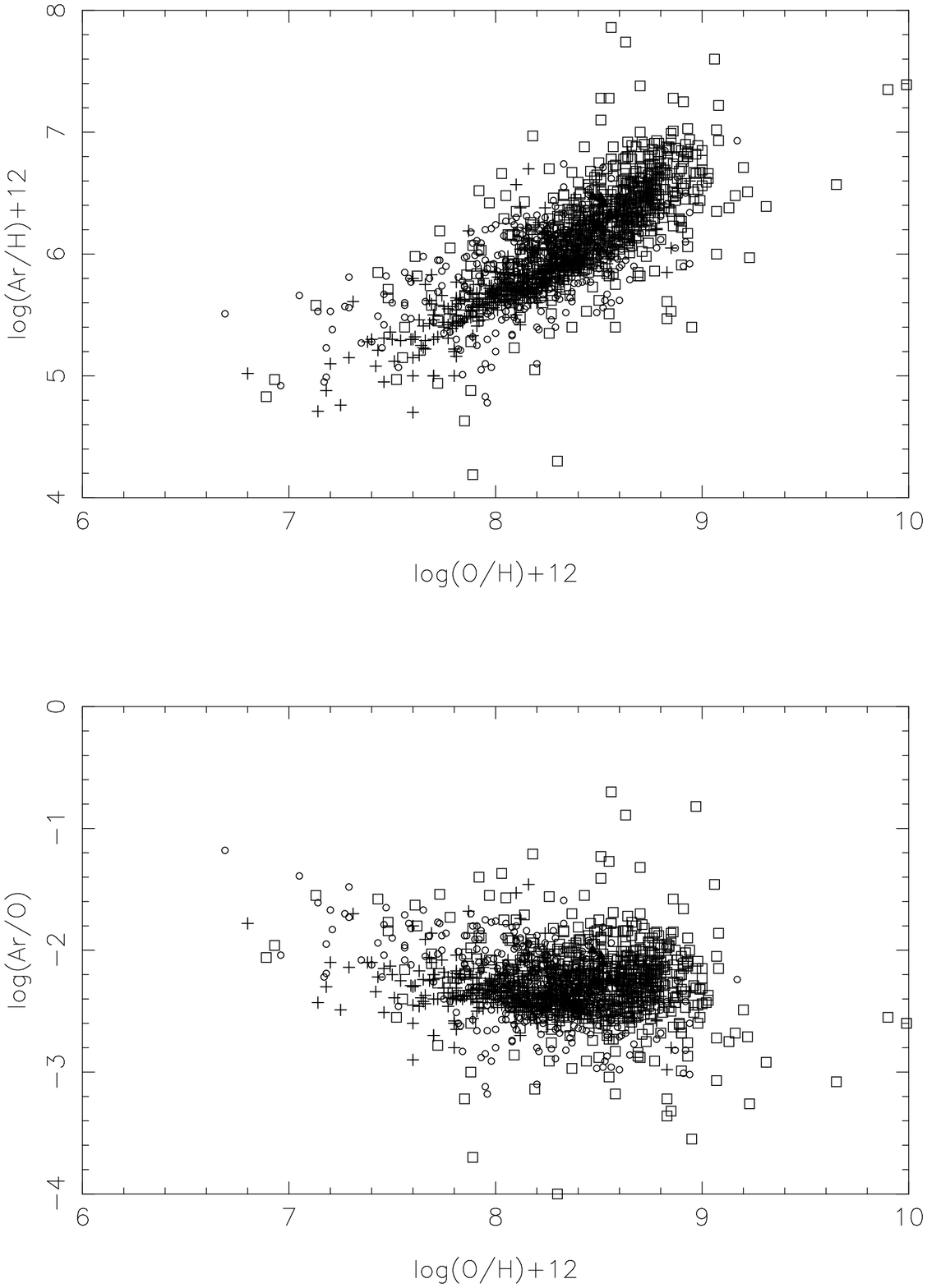}
   \caption{The same as Figure \ref{fig4} for argon.}
   \label{fig10}
   \end{figure}

For argon the results are similar compared to neon, but the dispersion is higher, comparable to S/H, as can 
be seen from Table~4 and Figures \ref{fig8}, \ref{fig9}, and \ref{fig10}, again with the same symbols as 
in Figures \ref{fig2} to \ref{fig4}.  The comparison with HII regions suffers from the lack of data for this 
element, especially for the Milky Way, except when external galaxies are included, as shown in 
Figure \ref{fig10}. The inclusion of BCG and ELG clearly improves the correlation, showing that the correlation 
defined at higher metallicities for the Milky Way still holds for lower and intermediate oxygen abundances 
$(\log {\rm O/H} + 12 < 8.5$). It should be noticed here that although the Ar/H dispersion is higher for PN 
compared with the Ne abundances, in the case of HII regions it is essentially the same as for Ne/H and S/H, 
up to 0.2 dex, similar to the uncertainty value. Therefore, the HII regions are clearly more homogeneous than 
the PN, which reflects their very low ages, roughly a few million years. Clearly, within such a short time 
bracket the average interstellar abundances are not expected to change appreciably. The relatively small 
dispersion of the HII region data for external galaxies can be understood in terms of the different 
metallicities of the Local Group objects. For PN, the dispersion in the data is considerably higher. As in 
the case of Ne, there is some evidence of a reduced oxygen abundance at lower metallicities.

\subsection{Elements produced by the PN  progenitor stars}
\label{subsection32}

In this work, we have stressed the elements that are not expected to be substantially produced by the PN 
progenitor stars, and we will present only a general outline of the elements that are strongly affected by 
the progenitor star evolution, such as He and N.  Carbon abundances are also modified by the stellar
evolution, but very few reliable carbon abundances for photoionized nebulae are presently available,
so that we will not consider this element in this work. The abundances of these elements are particularly 
modified by the dredge-up processes that occur in the intermediate mass stars. 

\begin{figure}
   \centering
   \includegraphics[angle=-90, width=5.5cm]{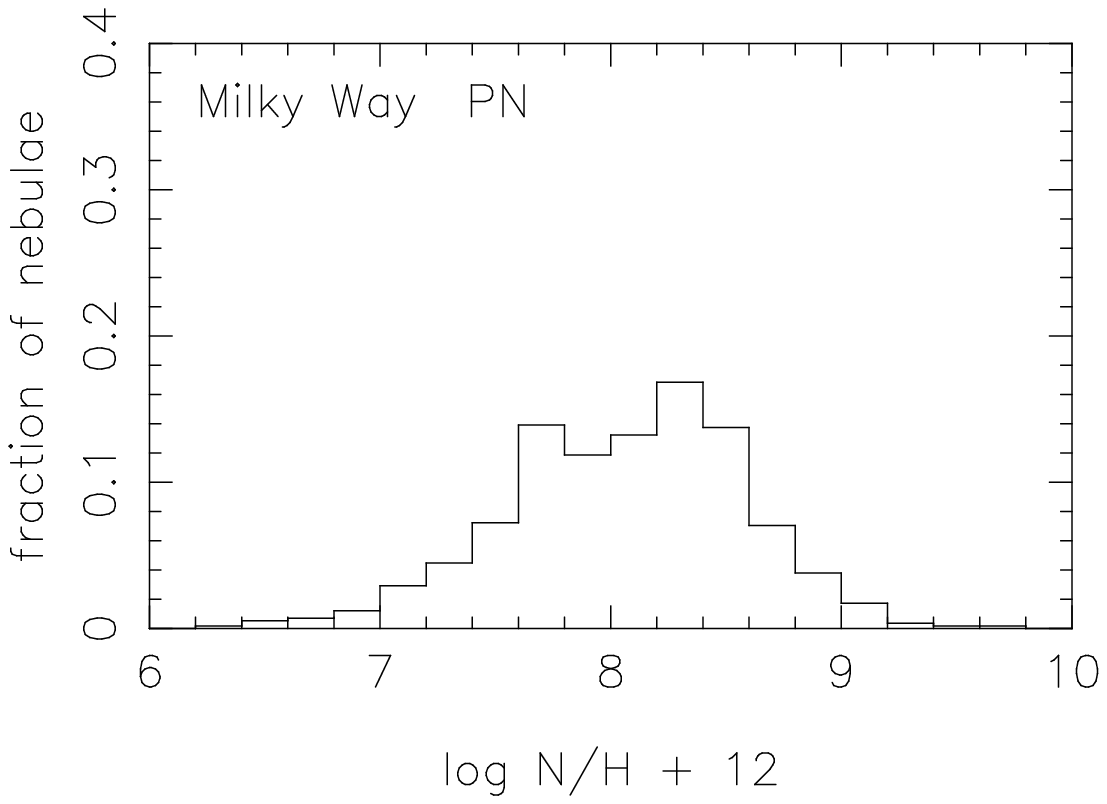}
   \includegraphics[angle=-90, width=5.5cm]{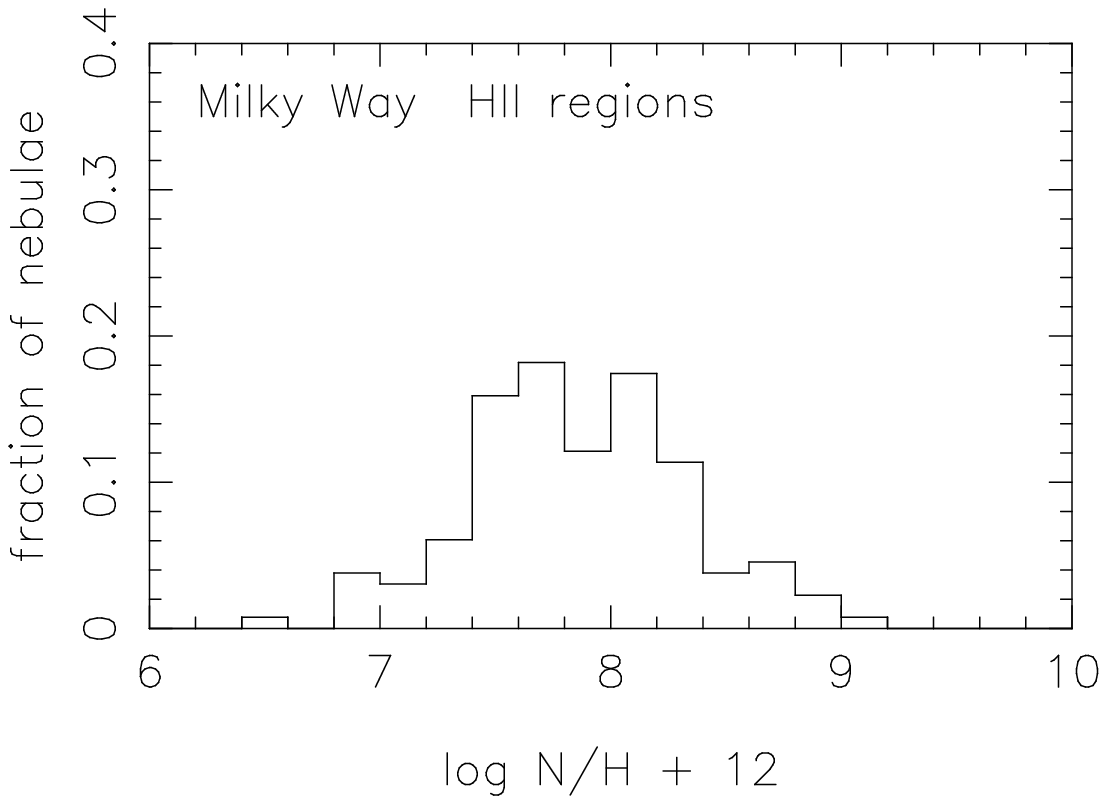}
   \includegraphics[angle=-90, width=5.5cm]{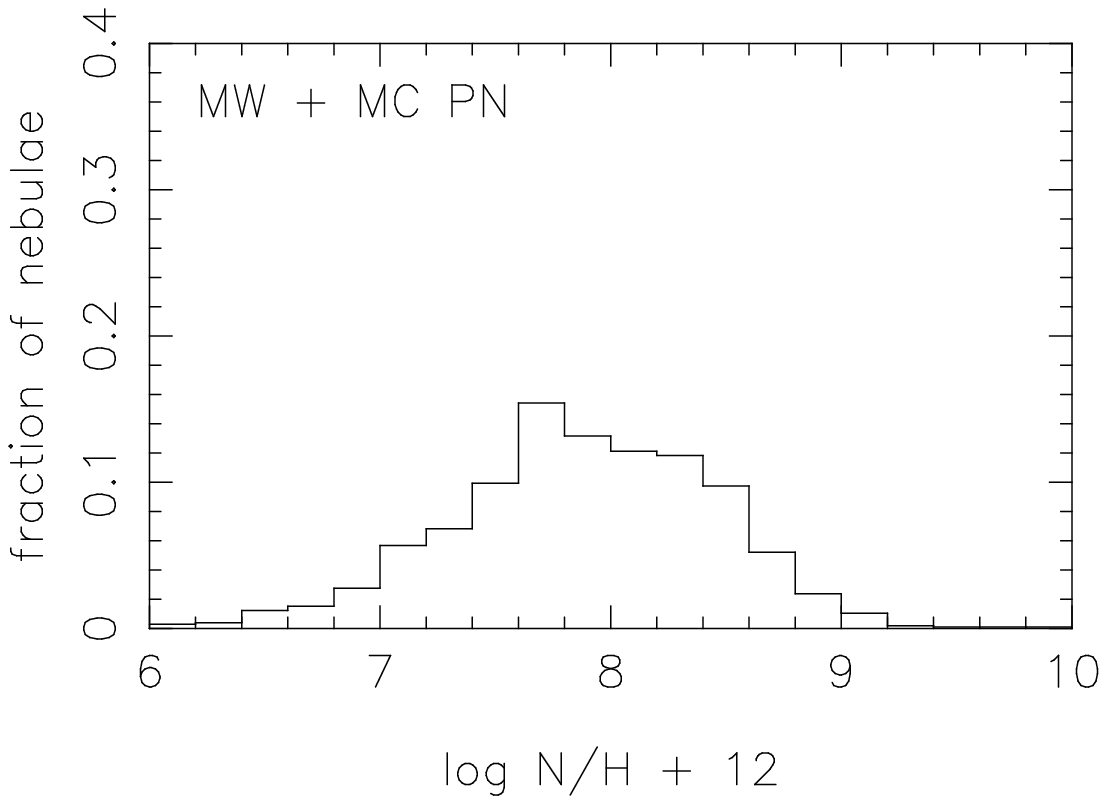}
   \includegraphics[angle=-90, width=5.5cm]{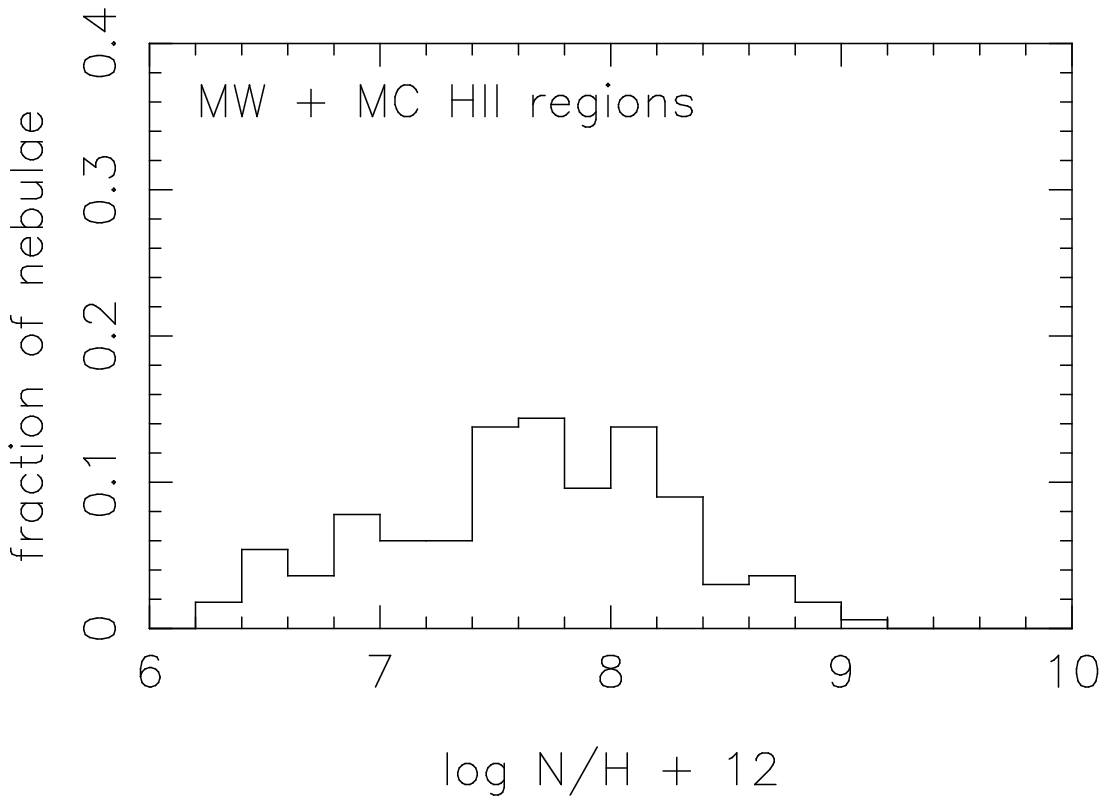}
   \includegraphics[angle=-90, width=5.5cm]{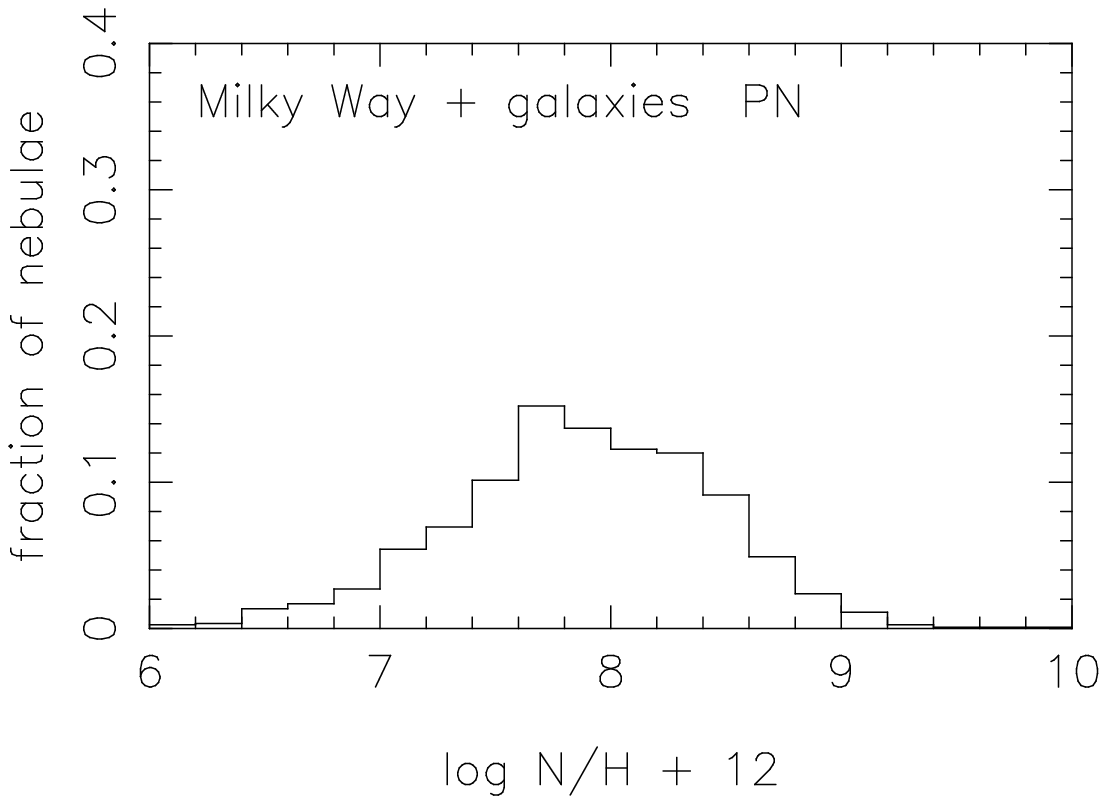}
   \includegraphics[angle=-90, width=5.5cm]{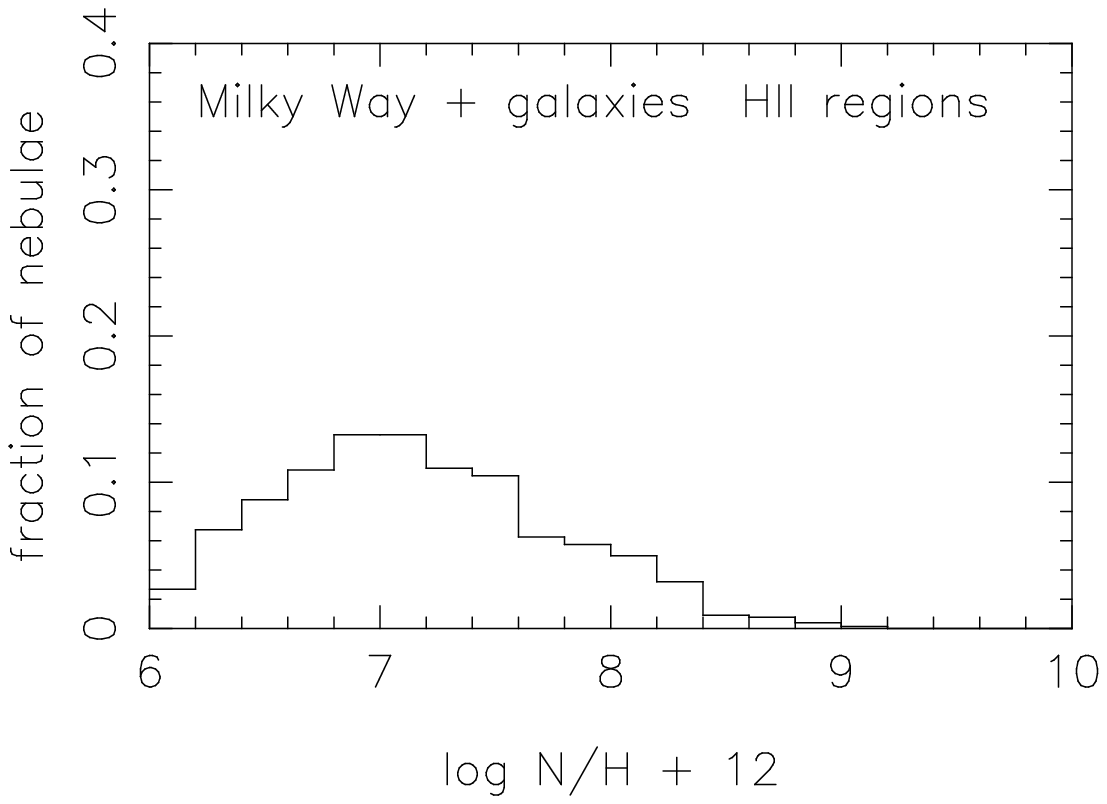}
   \caption{Histograms of the N/H abudances in PN and HII regions. top: Milky Way,
   middle: Milky Way and Magellanic Clouds, bottom: Milky Way and all external galaxies.}
   \label{fig11}
   \end{figure}

Histograms of the N/H abundances for PN and HII regions in our sample are shown in Figure \ref{fig11},
which can be directly compared with Figure \ref{fig1}. The PN distribution is similar is the three
cases shown, and peaks around $\log ({\rm N/H}) = 7.6 \ {\rm to} \ 8.4$, while for HII regions
the same is observed for the Milky Way and Magellanic Clouds, but the inclusion of external galaxies
(as well as BCG and ELG) shifts the maximum downwards by about 0.5 dex. The main difference between O/H 
and N/H as shown in Figures \ref{fig1} and \ref{fig11} is that the nitrogen abundances extends to lower 
metallicities for HII regions compared with planetary nebulae, which reflects the N production during  
the evolution of the PN progenitor stars.

The main elements that are produced by the PN progenitor stars are He and N, for which there are generally 
good quality data for planetary nebulae and, to a lesser extent, for HII regions. Also, He determinations in 
HII regions  are frequently affected by the presence of neutral helium, so that in this case the derived 
abundances are usually lower limits, which makes their analysis more difficult. In this work we have adopted 
a lower limit of He/H = 0.03 in order to avoid objects with an important fraction of neutral He.
 
   \begin{figure}
   \centering
   \includegraphics[angle=0, width=13.0cm]{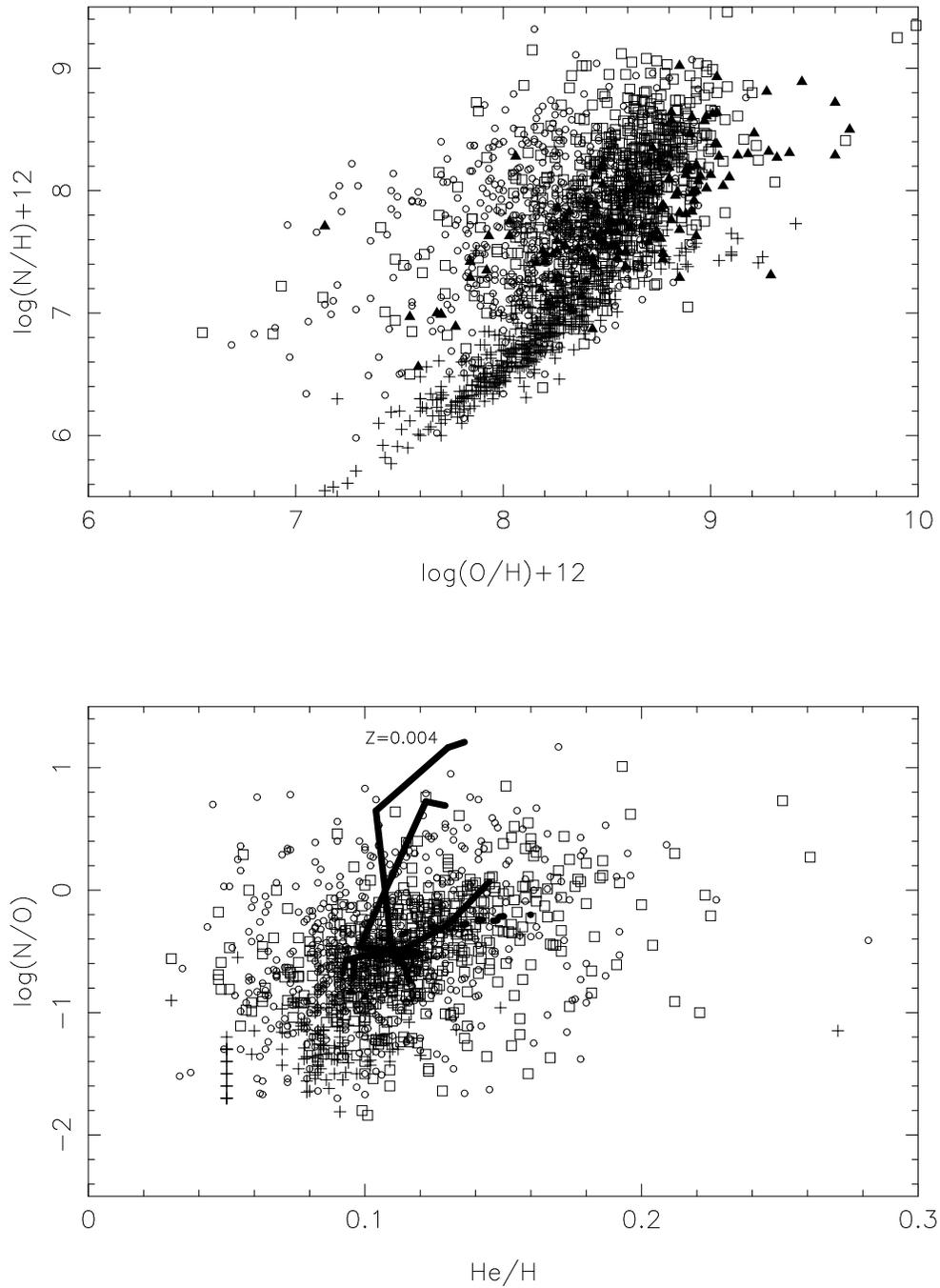}
   \caption{Nitrogen abundances as functions of O/H (top) and  He/H (bottom) for the Milky Way and external 
    galaxies. MW PN (squares), MW HII regions (triangles), external PN (circles), external HII 
    regions (crosses). In the bottom figure the thick solid lines represent models by Karakas (2010) for
    thermally pulsating AGB stars with Z = 0.004, as indicated at the top of the figure, and also for Z = 0.008 
    and Z = 0.02. The lower thick interrupted lines are models by Marigo et al. (2003) for Z = 0.019.}
   \label{fig12}
   \end{figure}

Figure \ref{fig12} shows the N/H and N/O ratios as functions of O/H and He/H, respectively,  for the case 
where all objects are considered, namely, the Milky Way, the Magellanic Clouds, and external galaxies, 
keeping the same symbols as in Figure \ref{fig4}. This figure can be compared with the bottom Figures
\ref{fig4}, \ref{fig7}, and \ref{fig10} for Ne, S, and Ar, respectively. The most striking result is that, 
as expected, PN show an increase in both N and He compared to most HII regions in the sample.  The average
dispersions of the nitrogen data are also shown in Table 4, and it can be seen that they are much larger
than in the case of Ne, S, and Ar, both for PN and HII regions. In other words, a larger dispersion is also
observed for HII regions, so that part of their nitrogen is probably secondary. The excess nitrogen in PN is 
essentially produced by their progenitor stars, in a strong contrast with the remaining elements studied so 
far. The N/H ratio is roughly correlated with O/H for HII regions (Figure \ref{fig12} top), but most PN are 
located above the average curve defined by the HII regions, again confirming the excess N in planetary nebulae.

Considering now Figure \ref{fig12} (bottom), it can be seen that there is a general trend for N/O with He/H for 
galactic as well as external planetary nebulae, but the dispersion is much higher than in the case of the previous 
plots, so that no definite correlation can be obtained, in agreement with similar conclusions by Richer \& McCall 
(\citeyear{richer2008}). Such a result clearly reflects the fact that the nitrogen abundances measured in 
planetary nebulae include both the pristine nitrogen plus the contribution from the dredge up processes that 
affect the red giant progenitor stars. Similar trends have also been recently discussed by Garc\'\i a-Hern\'andez 
et al. (\citeyear{garcia2016}), in a comparison of double-chemistry and oxygen-chemistry PN.  It can also be seen 
from Figure~\ref{fig12} (bottom) that there are no important differences between galactic and external objects, and 
the majority of PN have abundances close to solar, a result similar to the one obtained by Pottasch \& Bernard-Salas 
(\citeyear{pottasch2010}) for a smaller sample.

\begin{figure}
   \centering
   \includegraphics[angle=0, width=10.0cm]{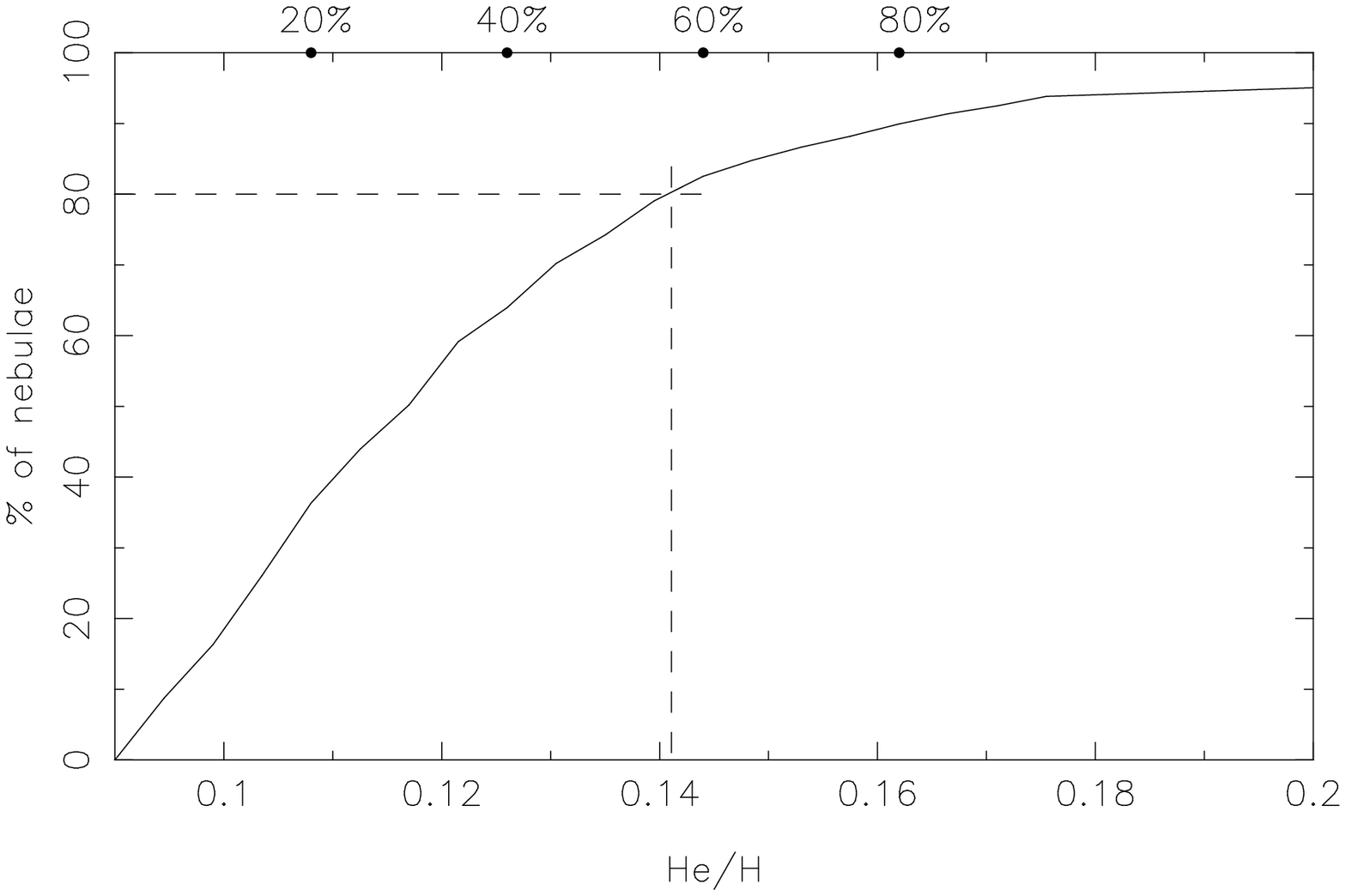}
   \includegraphics[angle=0, width=10.0cm]{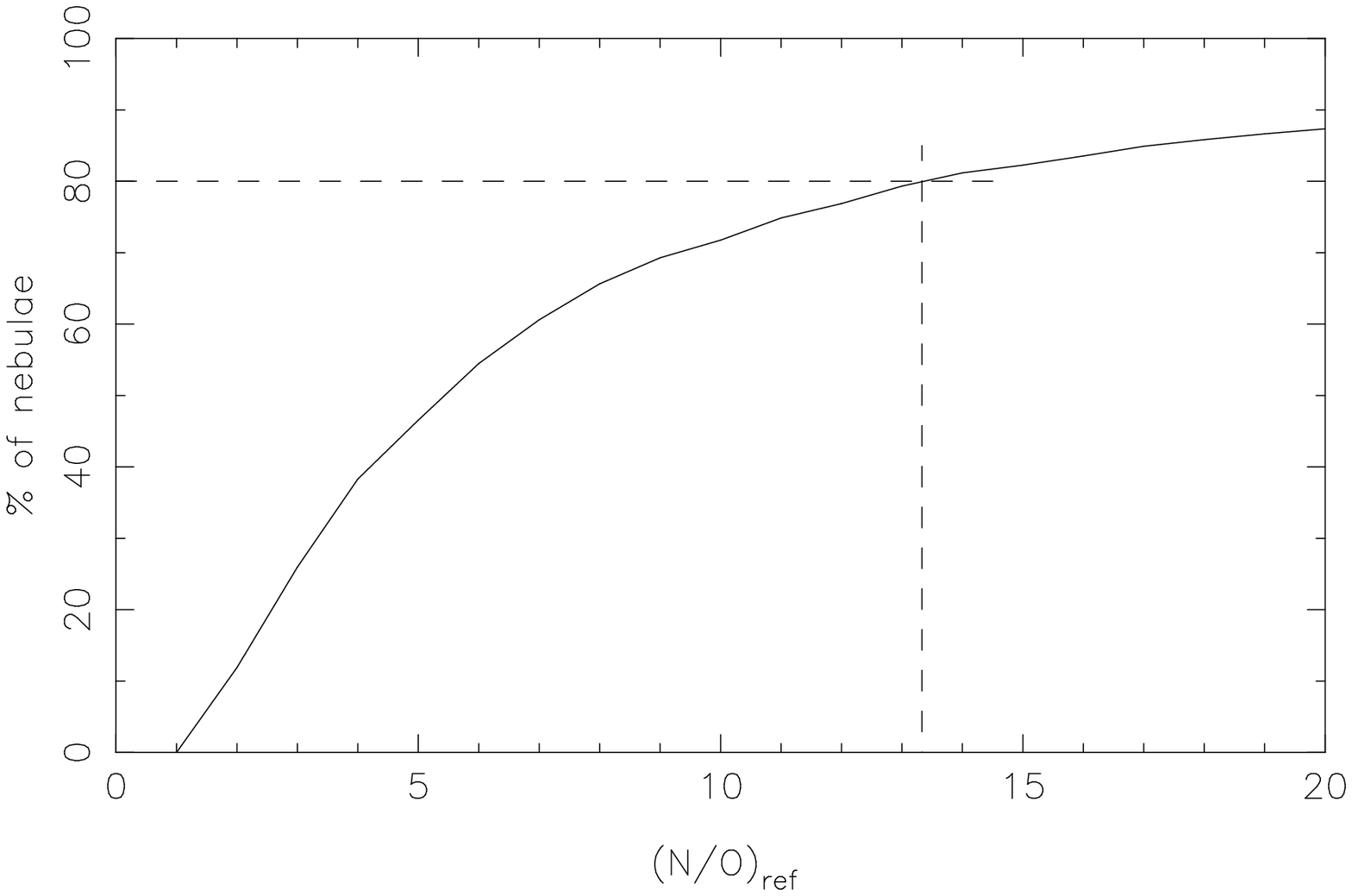}
   \caption{Top: Cumulative distribution of PN as a function of the He/H abundance. About
   80\% of the nebulae have He/H lower than 0.141, as indicated by the dashed lines, which 
   corresponds to a 57\%  enrichment relative to the reference value (He/H = 0.09). Bottom: 
   Cumulative distribution of PN as a function of the N/O enrichment ratio. About 80\% of the 
   nebulae have an enrichment ratio of a factor 13.3 relative to the reference value 
   (log N/O = --1.2), as indicated by the dashed lines.}
   \label{fig13}
   \end{figure}

The N/O ratio  shows some tendency to increase with the He abundances, indicating that both elements are produced 
the previous phases of the stellar evolution. Adopting a pregalactic He abundance by mass of about $Y = 0.255$ (cf. 
Izotov et al. \citeyear{izotov2014}) which corresponds to approximately He/H = 0.09, we see from Figure \ref{fig12} 
(bottom) that the PN enrichment is considerably larger than for HII regions, and that the difference between 
local and external PN is negligible. The cumulative distribution of the PN as a function of the He/H abundance 
is shown in Figure \ref{fig13} (top). It can be seen that about 80\% of the PN with He excess have abundances
up to He/H $\simeq 0.141$, which is about 57\% higher than the pregalactic value. This can be compared with an 
amount of 50\% derived by Richer \& McCall (\citeyear{richer2008}) from a smaller sample. More recently, 
Lattanzio \& Karakas (\citeyear{lattanzio2016}, see also Karakas \& Lattanzio \citeyear{karakas2014}) suggested 
an increase of about 38\% in the helium content by mass from the second dredge-up process in AGB stars, which 
would lead to an increase of about 60\% in the He abundance by number  of atoms, in excellent agreement with 
the results shown in Figure \ref{fig13}. 

Also from Figure \ref{fig12} we can have an idea of the amount of nitrogen produced by the progenitor stars. 
Adopting as limit for primary nitrogen the amount produced by type II supernovae (Izotov et al. \citeyear{izotov2006}), 
corresponding to approximately $\log {\rm N/O} = -1.6$, it can be seen that practically all PN are located high 
above this threshold. Considering the expected secondary nitrogen enrichment, which corresponds to about 
$\log {\rm N/O} = -1.2$,  Figure~13 (bottom) shows that about 80\% of the PN present an enrichment ratio up to  
factor of 13.3, comparable with the factor of 10 found by Richer \& McCall (\citeyear{richer2016}).

In Figure \ref{fig12} (bottom) we include a comparison with some recent theoretical models. The thick solid lines 
are from theoretical models by Karakas (\citeyear{karakas2010}, see also Karakas \& Lattanzio 
\citeyear{karakas2007} and Garc\'\i a-Hern\'andez \& G\'orny \citeyear{garcia-gorny}), with  $Z = 0.004,\ 0.008,$ 
\ and $0.02$, while the thick interrupted lines represent models by Marigo et al. (\citeyear{marigo2003}) 
with $Z = 0.019$.  We have used straight lines to link the model results effectively obtained by 
Marigo et al. (\citeyear{marigo2003}) and Karakas (\citeyear{karakas2010}). These are synthetic 
evolutionary models for thermally-pulsing AGB stars with initial masses of 1 to 6 $M_\odot$, in which 
up to three dredge-up episodes occur, apart from hot-bottom processes (HBB) for the most massive objects.  
According to these models, progenitors having 0.9 to 4 $M_\odot$ and solar composition can explain the 
\lq\lq normal\rq\rq\ abundances, He/H $< 0.15$, while for objects with higher enhancements (He/H $ > 0.15$), 
masses of 4 to 5 $M_\odot$ are needed, plus an efficient HBB. For intermediate mass stars, agreement 
with theoretical models is fair, but abundance determinations should be improved and expanded. Recent 
models by Pignatari et al. (\citeyear{pignatari}) with $Z = 0.01$ and 0.02 are also consistent with 
these results, as can be seen in the discussion by Delgado-Inglada et al. (\citeyear{delgado2015}). 
This paper also includes a discussion on the effects of different ICFs on the N and He production in the 
PN progenitor stars, as applied to a smaller sample than considered here.  

\section{Conclusions}
\label{section4}

The main conclusions of this paper are as follows:

\bigskip\noindent
$\bullet$\ PN abundances of O, Ne, S, and Ar show good correlations, indicating that Ne, S, and Ar vary in lockstep 
with O. The correlations also apply to HII regions, BCG and ELG, but the dispersion is considerably
lower for these objects. 

\bigskip\noindent
$\bullet$\ No important contribution to the Ne abundances from their progenitor stars is found for PN, so that any 
such contribution is probably lower than the current uncertainties in the abundance determination.

\bigskip\noindent
$\bullet$\ Argon presents similar results as neon, with a somewhat higher dispersion, probably due to the 
weakness of the Ar lines. 

\bigskip\noindent
$\bullet$\ Sulphur abundances in Milky Way PN may present the sulphur anomaly, that is, lower abundances than 
expected at a given metallicity. This is  probably due to incorrect ICFs (ionization correction factors), but 
is not apparent in the bulk of Local Group objects, for which a particularly extended metallicity range can be 
observed. The anomaly is not observed in MC PN, which show instead a larger dispersion than in the 
Milky Way. The dispersion observed in PN is probably real, and partially reflects the different ages of the 
progenitor  stars. The variety of the sources considered may also contribute to the observed dispersion,
although an effort has been made to consider only works with similar approaches in the abundance
determination.

\bigskip\noindent
$\bullet$\  All correlations for HII regions show smaller dispersions, as expected, a result that is not 
affected by the inclusion of either BCG or ELG. The same trends are observed in both types of photoionized 
nebulae. 

\bigskip\noindent
$\bullet$\  Concerning planetary nebulae, it is interesting to notice that the average dispersion of the relations 
involving Ne, Ar, and S with oxygen  is similar to the more homogeneous sample by Henry and co-workers, 
as can be seen for example in Milingo et al. (\citeyear{milingo}) (Here the remarks made on sulphur data 
are also applied).  This is very interesting, considering the fact that the Local Group galaxies have 
somewhat different metallicities, and possibly some differences in their chemical evolution processes. 
A similar result was obtained by Richer \& McCall (\citeyear{richer2008}) for neon. Also for neon and 
sulphur, results by Magrini et al. (\citeyear{magrini2009}) for M33 agree with the conclusions above. 
More recently, Richer \& McCall (\citeyear{richer2016}) pointed out that the relation of oxygen and 
neon abundances is similar to the star-forming galaxies, which again suggests that the PN progenitor 
stars do not appreciably change the Ne abundances or the possible changes keep the original abundance 
ratio, implying progenitor masses of about 2 solar masses or less.

\bigskip\noindent
$\bullet$\  The same mass interval is obtained from the comparison of the nitrogen abundances in PN with 
theoretical models of AGB stars, as PN have higher N and He abundances, and are located at the top right corner 
of the N/O $\times$ He/H plot. Therefore, these objects have originated from intermediate mass stars with 
masses typically under 2 $M_\odot$, except for the high He nebulae, whose progenitor stars are closer 
to the high mass bracket, around  4 $M_\odot$.

\bigskip\noindent
$\bullet$\  A comparison of the N and He abundances in PN and HII regions shows that the former present an 
enrichment of  these elements up to a factor of 13 for nitrogen and up  to 57\% for He, which agrees with some 
recent observational determinations as well as theoretical models.

\bigskip\noindent
Acknowledgements. This work was partially supported by FAPESP and CNPq.

\end{document}